\documentclass{article}

\usepackage{arxiv}

\usepackage[utf8]{inputenc} 
\usepackage[T1]{fontenc}    
\usepackage{hyperref}       
\usepackage{url}            
\usepackage{booktabs}       
\usepackage{amsfonts}       
\usepackage{nicefrac}       
\usepackage{microtype}      
\usepackage{graphicx}
\usepackage{amssymb}
\usepackage{amsmath}
\usepackage{lineno}
\usepackage{hyperref}
\usepackage{listings}
\usepackage{subfigure}
\usepackage{comment}
\usepackage{enumitem}
\usepackage[dvipsnames]{xcolor}

\title{Stochastic Multi-Dimensional Deconvolution}

\author{
  Matteo Ravasi \\
  KAUST\\
  Thuwal, Kingdom of Saudi Arabia \\
  \texttt{matteo.ravasi@kaust.edu.sa} \\
   \And
 Tamil Selvan \\
  Anna University \\
  Chennai, India \\
  \And
 Nick Luiken \\
  KAUST\\
  Thuwal, Kingdom of Saudi Arabia\\
  \texttt{nicolaas.luiken@kaust.edu.sa} \\
  }
  
\begin{document}

\chead{Stochastic Multi-Dimensional Deconvolution}

\maketitle

\begin{abstract}
  Seismic datasets contain valuable information that originate from areas of interest in the subsurface; such seismic reflections are however inevitably contaminated by other events created by waves reverberating in the overburden. Multi-Dimensional Deconvolution (MDD) is a powerful technique used at various stages of the seismic processing sequence to create ideal datasets deprived of such overburden effects. Whilst the underlying forward problem is well defined for a single source, a successful inversion of the MDD equations requires availability of a large number of sources alongside prior information introduced in the form of physical preconditioners (e.g., reciprocity). In this work, we reinterpret the cost function of time-domain MDD as a finite-sum functional, and solve the associated inverse problem by means of stochastic gradient descent algorithms, where gradients are computed using a small subset of randomly selected sources. Through synthetic and field data examples, the proposed method is shown to converge more stably than the conventional approach based on full gradients. Stochastic MDD represents a novel, efficient, and robust strategy to deconvolve seismic wavefields in a multi-dimensional fashion.
\end{abstract}

\section*{Introduction}
Multi-Dimensional Deconvolution (MDD) is a versatile technique used in seismic processing and imaging to suppress the effect of an unwanted overburden; this is simply achieved by deconvolving the up- and down-going components of the recorded seismic data at a certain datum of interest. In this context, the word \textit{overburden} may simply refer to the free-surface when data are acquired via a streamer or ocean-bottom cable (OBC) acquisition system, or a portion of the Earth when recordings are available at a certain depth within the subsurface in either a horizontal borehole or estimated by means of model-based (e.g., \cite{biondi2018, guo2020, li2021}) or data-driven (e.g., \cite{Wapenaar2014, vanderneut2015, ravasi2017}) redatuming techniques. More recently, the applicability of MDD has been also extended to laboratory acoustic measurements with the aim of removing from the recorded data unwanted reflections originated at the boundaries of the experimental setup \cite{Li2021b}. Such a finding may have also broader implications in other domains of science such as non-destructive testing or medical imaging; in both cases, acoustic waves originated from discontinuities inside a probed body are in fact polluted by strong reverberations from the rigid boundary between such a body and the fluid (i.e., air or water) that encompasses it.

The original theory of MDD dates back to the seminal work of \cite{amundsen2001}. However, for a long time, the geophysical community has only embraced a 1D approximation of such a theory as it leads to an efficient element-wise, stabilized division in the frequency-wavenumber ($f-k$) or linear Radon ($\tau-p$) domains \cite{lokshtanov2000, wang2010}, sometimes referred to as to as Up/Down Deconvolution (UDD). Whilst such an approximation is valid for a horizontally invariant overburden, its accuracy deteriorates significantly when this assumption is not met. A special case is represented by a dipping seabed for the free-surface multiple attenuation scenario, as the only interface in the overburden is represented by the sea-air interface, which is not aligned with the seafloor. Within this context, \cite{boiero2020} have recently studied the validity of the UDD method against the more accurate MDD method, and concluded that the latter must be used for seabed dips beyond $1^\circ$. Similarly, when the purpose of MDD is that of suppressing the effect of a highly complex overburden within an imaging context, the 1D approximation is never valid. This explains the inability of single-channel deconvolution between the source (down-going) and receiver (up-going) wavefields to handle cross-talk artefacts when jointly imaging primaries and multiples \cite{muijs2007, lu2015}.

Despite the theoretical superiority of MDD over UDD, the arguments that Multi-Dimensional Deconvolution is a severely ill-posed and difficult to stabilize inverse problem alongside its extreme computational cost have for a long time hindered the widespread adoption of such an algorithm in industrial settings. Early attempts to mitigate the ill-posed nature of such an inverse problem have been reported by \cite{minato} that employed Singular Value Decomposition (SVD) when solving MDD in the frequency domain. More recently, \cite{vanderNeut2013} proposed a time-domain formulation of the problem to naturally regularize the inverse problem and used sparsity-promoting inversion to further mitigate the effect of noise in the input data. Along similar lines, \cite{vanderneut2017} and \cite{luiken2020} proposed to introduce a preconditioner to enforce reciprocity in the coveted target response. \cite{vargas2021b} combined three physics-based preconditioners (i.e., causality, reciprocity, and frequency-wavenumber locality) alongside providing a comprehensive analysis of their individual and combined impact on the solution of MDD in the presence of noisy input data retrieved by means of the so-called Scattering-Rayleigh Marchenko redatuming method (SRM - \cite{vargas2021a}).

In this work, after reinterpreting the time-domain MDD cost function as a finite-sum functional, we propose to solve the associated inverse problem using stochastic gradient descent algorithms. This is primarily motivated by the need to reduce the overall computational cost of MDD and supported by the physical argument that nearby sources are likely to provide redundant contributions to the gradients that drive the time-domain MDD inversion. As a by-product of this new algorithmic choice, we observe an overall increased stability in the convergence properties of MDD, making it less reliant on the choice of the maximum number of iterations or stopping criterion for the gradient-based solver of choice. The proposed algorithm is applied to three synthetic datasets and the Volve field dataset. As far as synthetic data are concerned, the first and third examples focus on the demultiple problem of up- and down-going data from an OBC acquisition geometry along a non-flat seabed. The second example considers instead subsurface-to-surface wavefields obtained by means of SRM redatuming, which requires us to deal with a severely ill-posed MDD process \cite{vargas2021b}. This paper is structured as follows. First, the theory of stochastic MDD is presented. This is followed by a section containing a set of numerical examples. We conclude by discussing the benefits of the proposed methodology and its outstanding challenges.

\section*{Theory}
The up- ($p^-(\textbf{x}_{VS}, \textbf{x}_{S}, f)$) and down-going ($p^+(\textbf{x}_{R}, \textbf{x}_{S}, f)$) components of the seismic wavefield from a source $\textbf{x}_S$ to a line of receivers $\textbf{x}_R$ at a given datum $\partial \mathbb{D}$ and another receiver $\textbf{x}_{VS}$ at any location below $\partial \mathbb{D}$ are linked to the local reflection response $R(\textbf{x}_{R}, \textbf{x}_{VS}, f)$ via the following Multi-Dimensional Convolution (MDC) integral in the frequency domain \cite{amundsen2001, wapenaar2011}:
\begin{equation}
\label{eq:mddintegral}
p^-(\textbf{x}_{VS}, \textbf{x}_{S}, f) = \int_{\partial \mathbb{D}} p^+(\textbf{x}_{R}, \textbf{x}_{S}, f) R(\textbf{x}_{R}, \textbf{x}_{VS}, f) d\textbf{x}_R
\end{equation}
where $f$ is the frequency. In this context, the \textit{local reflection response} is intended as the seismic response of an ideal medium that is equivalent to the original medium below the datum $\partial \mathbb{D}$ and homogeneous above, therefore deprived of any interaction with the overburden. The process of inverting equation \ref{eq:mddintegral} is usually referred to in the literature as Multi-Dimensional Deconvolution (MDD). 

\begin{figure}
  \centering
  \includegraphics[width=0.5\textwidth]{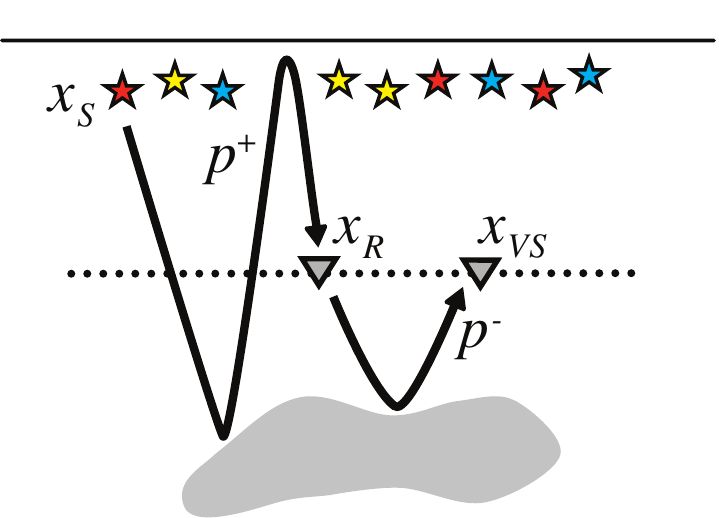}
  \caption{Schematic representation of the stochastic MDD algorithm. Red, yellow, and cyan sources represent three different batches with $N_{s,batch}=3$. Grey triangles indicate a receiver and a virtual source along a sample raypath composed of a down-going ($p^+$) and up-going ($p^-$) event linked via equation \ref{eq:mddintegral}.}
  \label{fig:mddschematic}
\end{figure}

Whilst the forward problem is well-posed for a single source, the solution of equation \ref{eq:mddintegral} is a severely ill-posed inverse problem (see, for example, \cite{minato}). It therefore requires availability of multiple sources, which ideally should be equal to or exceed the number of receivers along the datum $\partial \mathbb{D}$ (i.e., $N_s \ge N_r$). Moreover, it is common practice to consider multiple virtual sources $\textbf{x}_{VS}$ placed along the same datum $\partial \mathbb{D}$ and express equation \ref{eq:mddintegral} in a compact matrix-matrix notation for each frequency $f$:
\begin{equation}
\label{eq:mdd}
\textbf{P}_f^- = \textbf{P}_f^+ \textbf{R}_f
\end{equation}
where $\textbf{P}_f^-$ is a matrix of size $\left[ N_s \times N_{vs} \right]$, $\textbf{P}_f^+$ is a matrix of size $\left[ N_s \times N_{r} \right]$, and $\textbf{R}_f$ is a matrix of size $\left[ N_r \times N_{vs} \right]$. This system of equations can be solved with direct methods \cite{wapenaar2011, minato} or iterative Block-Krylov solvers \cite{luiken2019}. Alternatively, a time-domain formulation can be written as \cite{vargas2021b}:
\begin{equation}
\label{eq:mddtinv}
\textbf{r}_t = \underset{\mathbf{r}_t \in \mathbb{C}} {\mathrm{argmin}}  \frac{1}{2}||\textbf{p}_t^- - \textbf{P}_t^+ \textbf{r}_t||_2^2
\end{equation}
where $\textbf{p}_t^-$ is a vector of size $\left[ N_s N_{vs}N_t \times 1 \right]$ and $\textbf{r}_t$ is a vector of size $\left[ N_r N_{vs} N_t \times 1 \right]$ obtained by stacking traces for all available pairs of sources (or receivers) and virtual sources. Finally, $\textbf{P}_t^+$ is a linear operator applying Multi-Dimensional Convolution in the time domain as detailed in \cite{Ravasi2021}. Here the solution is forced to belong to a subspace of interest, $\mathbb{C}$, based on physical constraints (i.e., reciprocity and/or causality) that the wavefield is expected to satisfy. As discussed in detail in \cite{luiken2020, vargas2021b}, the constraints can be enforced by means of a chain of preconditioners ($\textbf{C} = \textbf{C}_r \textbf{C}_p$; $\textbf{C}_r$ for reciprocity and $\textbf{C}_c$ for causality):
\begin{equation}
\label{eq:mddtprecinv}
\textbf{z} = \underset{\mathbf{z}} {\mathrm{argmin}}  \frac{1}{2}||\textbf{p}_t^- - \textbf{P}_t^+ \textbf{C} \textbf{z}||_2^2
\end{equation}
and the solution can be finally obtained as $\textbf{r}_t = \textbf{C}\textbf{z}$. This unconstrained least-squares functional can be minimized by means of gradient-based iterative solvers (e.g., LSQR -- \cite{Paige1982}). For simplicity, in the remainder of the paper we will refer to the solution of equation \ref{eq:mddtprecinv} as \textit{full-gradient MDD}.

However, since the forward problem in equation \ref{eq:mddintegral} is valid for any of the available sources, the functional in equation \ref{eq:mddtinv} can be equivalently written as a finite-sum functional:
\begin{equation}
\label{eq:mddtfinite}
\textbf{r}_t = \underset{\mathbf{r}_t \in C} {\mathrm{argmin}} \frac{1}{2} \sum_{i_s=1}^{N_s} ||\textbf{p}_{t,i_s}^- - \textbf{P}_{t,i_s}^+ \textbf{r}_t||_2^2
\end{equation}
where $\textbf{p}_{t,s_i}^-$ is a vector of size $\left[ N_{vs}N_t \times 1 \right]$ that contains the up-going wavefield originated from the i-th source. Similarly, $\textbf{P}_{t,i_s}^+$ is a linear operator performing MDC with the down-going wavefield of the i-th source. Note that, a physical preconditioner may be added to this functional similar to that in equation \ref{eq:mddtprecinv}. 

Drawing upon the vast literature of stochastic gradient algorithms \cite{Robbins1951, Saad1998}, the inverse problem in equation \ref{eq:mddtfinite} can be conveniently solved in a mini-batch fashion. More specifically, rather than computing the exact gradient of equation \ref{eq:mddtfinite} at each iteration and using it to update the current $\textbf{r}_t$ vector, the available sources are grouped in batches of size $N_{s,batch}<N_{s}$ and the gradient of each batch $S_{batch}$ is computed as follows:
\begin{equation}
\label{eq:mddtgrad}
\textbf{g}_{t,S_{batch}} = -\sum_{i_s \in S_{batch}} \textbf{P}_{t,i_s}^{+H} (\textbf{p}_{t,i_s}^- - \textbf{P}_{t,i_s}^+ \textbf{r}_t)
\end{equation}
This approximated gradient is then used in the stochastic gradient descent (SGD) and the stochastic gradient descent with Nesterov Momentum (N-SGD) iterations, where the implementation of Nesterov momentum is based on \cite{Sutskever2013}. In the following, the solution of equation \ref{eq:mddtfinite} will be referred to as \textit{stochastic MDD}. Moreover, the word \textit{epoch} is used when referring to a set of iterations that utilize all of the available sources in a dataset, whilst \textit{iteration} or \textit{step} is used to indicate a single gradient step computed with a batch of sources. Finally, we note that by considering a subset of sources at each step of the stochastic solvers, the computational cost of applying the operator $\textbf{P}^+_{t, i_s}$ and its adjoint is also reduced by roughly the ratio of the overall number of sources over that of the sources belonging to a batch (see Appendix A for a more detailed analysis of the computational cost of the two MDD algorithms). In the following, when comparing the solutions of different algorithms, we will therefore assume that the cost associated with one epoch of a mini-batch algorithm is about the same as that of one iteration of a full-gradient algorithm. Therefore, to avoid confusion, we also use the word epoch when referring to a single gradient step in full-gradient MDD.

\section*{Examples}
\begin{figure*}[htb]
  \centering
  \includegraphics[width=0.99\textwidth]{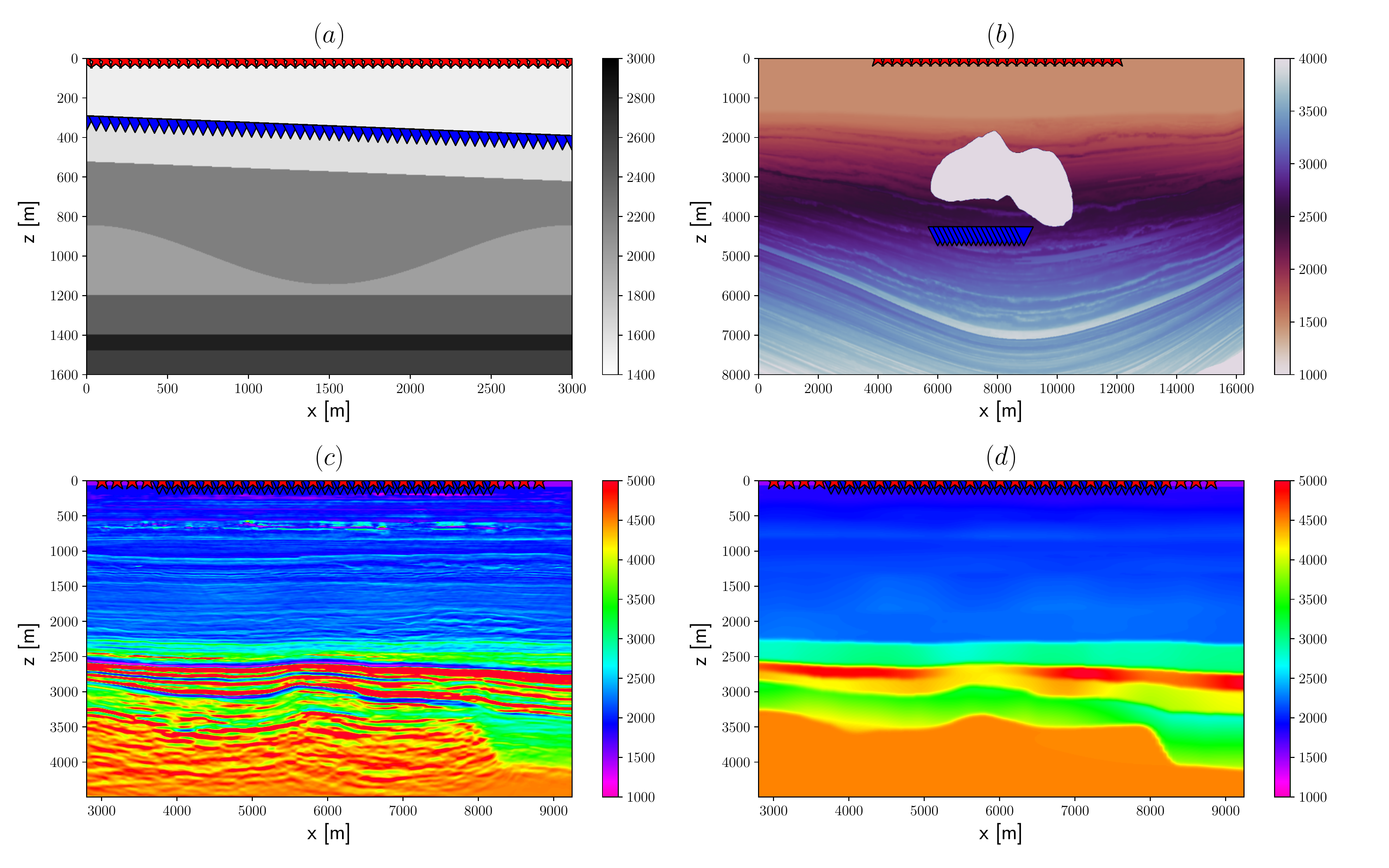}
  \caption{Velocity models and acquisition geometries (red stars indicate sources and blue triangles refer to receivers). a) Synthetic model with dipping seabed, b) Subsalt model, c) Synthetic Volve model, and d) Migration velocity model used to image the field Volve dataset.}
  \label{fig:models}
\end{figure*}

Four numerical examples are discussed in this section as depicted in Fig. \ref{fig:models}. The first and last two examples are similar in that they use seismic data after up/down decomposition as input to the MDD process. In these cases, we aim to create a virtual dataset that is free from surface-related multiples. Whilst the decomposition process is likely to introduce some leakage of the up-going component into the down-going and vice versa, we expect such errors to be minimal. In the second synthetic example instead, MDD is performed on up- and down-going wavefields that have been redatumed to a certain depth level by means of a data-driven redatuming process, namely SRM redatuming. This process inevitably introduces coherent artefacts in both the up- and down-going wavefields, as extensively discussed in \cite{vargas2021b}. The associated MDD problem is highly unstable and requires physical preconditioning to produce satisfactory results.

In all of the experiments, qualitative and quantitative assessment of the results produced by the full-gradient and stochastic MDD algorithms will be presented. As far as the latter is concerned, when dealing with synthetic data, the root mean-square error (RMSE) is used as metric:

\begin{equation}
\label{eq:rmse}
RMSE(\textbf{r}_t^{true}, \textbf{r}_t^{est}) = \sqrt{\Bigg |\Bigg |\frac{\textbf{r}_t^{true}}{max\{|\textbf{r}_t^{true}|\}} - \frac{\textbf{r}_t^{est}}{max\{|\textbf{r}_t^{est}|\}}\Bigg |\Bigg |_2^2}
\end{equation}
where $\textbf{r}_t^{true}$ is the true reflection response obtained by means of finite-difference modelling, $\textbf{r}_t^{est}$ is the estimated reflection response by means of MDD. Here, $||\textbf{x}||^2_2 = \sqrt{\sum_{i=1}^N x_i^2}$ is the squared Euclidean distance, and $max\{|\textbf{x}|\}=max\{|x_1|, |x_2|, ..., |x_N|\}$.

\subsection{Synthetic OBC Up/Down Deconvolution}
\begin{figure}
  \centering
  \includegraphics[width=0.7\textwidth]{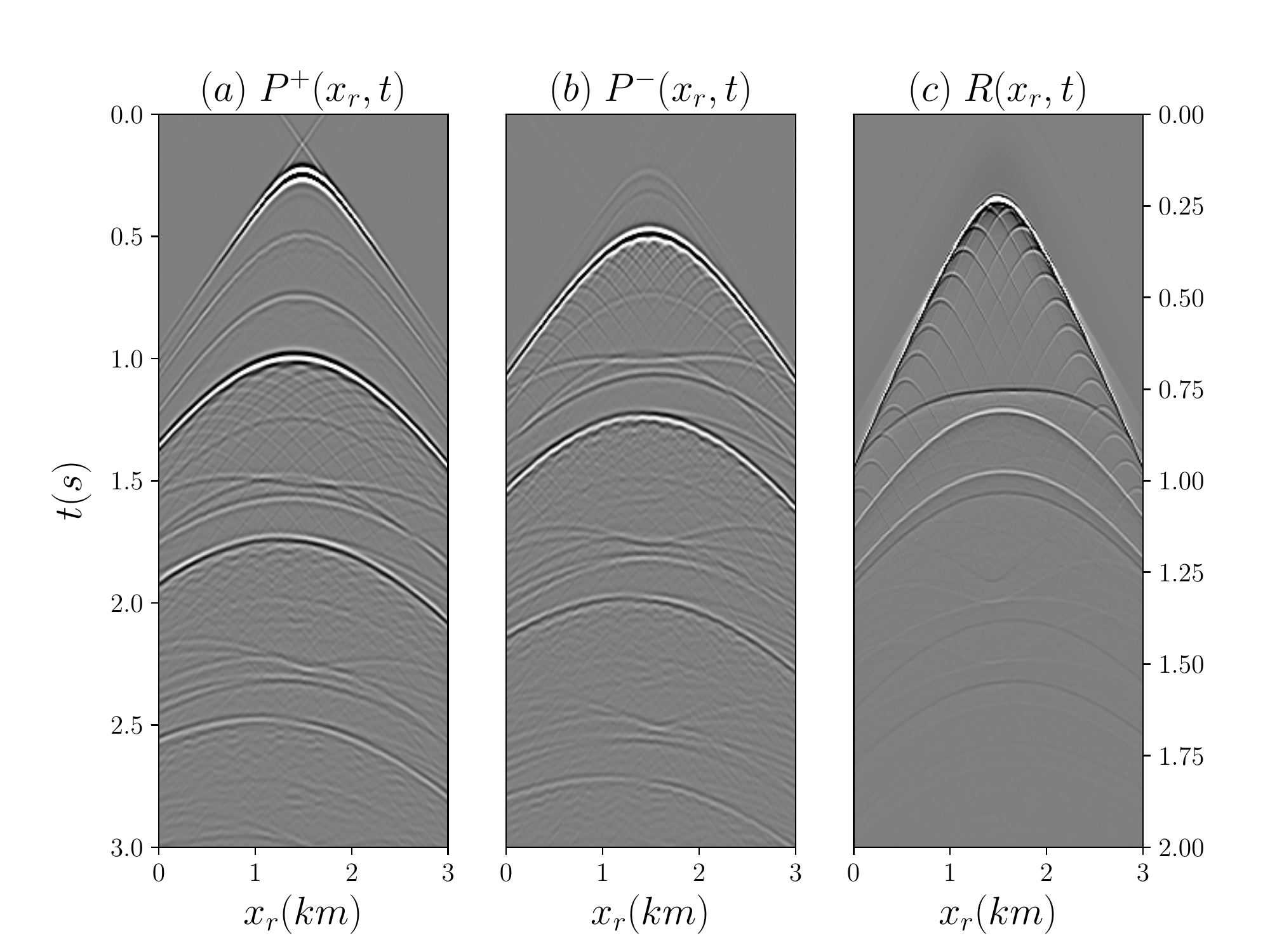}
  \caption{Common shot gather in the middle of the source array for the a) up-going wavefield, b) down-going wavefield, and c) ideal reflection response without free-surface effects.}
  \label{fig:obc_data}
\end{figure}

The first numerical example deals with the variable-density, variable-velocity model shown in Fig. \ref{fig:models}a. An ocean-bottom cable acquisition setup with receivers placed along a dipping seabed ($1.9^\circ$ dip) is used to create the dataset. This example is partially inspired by the work of \cite{boiero2020} that showed the importance of deconvolving the up- and down-going wavefields in a multi-dimensional fashion in the presence of a non-flat seafloor. Note that to ensure minimal error in the wavefield separation step, both the velocity and density of the seabed are chosen to be close to those of the water layer (i.e., soft seafloor). However, a hard dipping layer is introduced at around $600m$ depth; this reflector represents the main generator of strong free-surface multiples that we wish to tackle by means of MDD. A line of 201 sources at depth $z_s=10m$ with spatial sampling $dx_s=15m$ is used alongside a line of 201 receivers at varying depth ranging from $z_{r,min}=328m$ to $z_{r,max}=428m$  and spatial sampling $dx_r=15m$. A Ricker wavelet with central frequency $f_c=20Hz$ is used to model the pressure and vertical particle velocity data and a free-surface is added on top of the model. The dual-sensor data are subsequently separated into their up- and down-going pressure components in the ($\omega-k_x$) domain \cite{Wapenaar1998}, as shown in Fig. \ref{fig:obc_data}a and b. The ideal reflection response from sources and receivers at the seafloor and in the absence of both the free-surface and water column is also modelled and displayed in Fig. \ref{fig:obc_data}c.

\begin{figure}
  \centering
  \includegraphics[width=\textwidth]{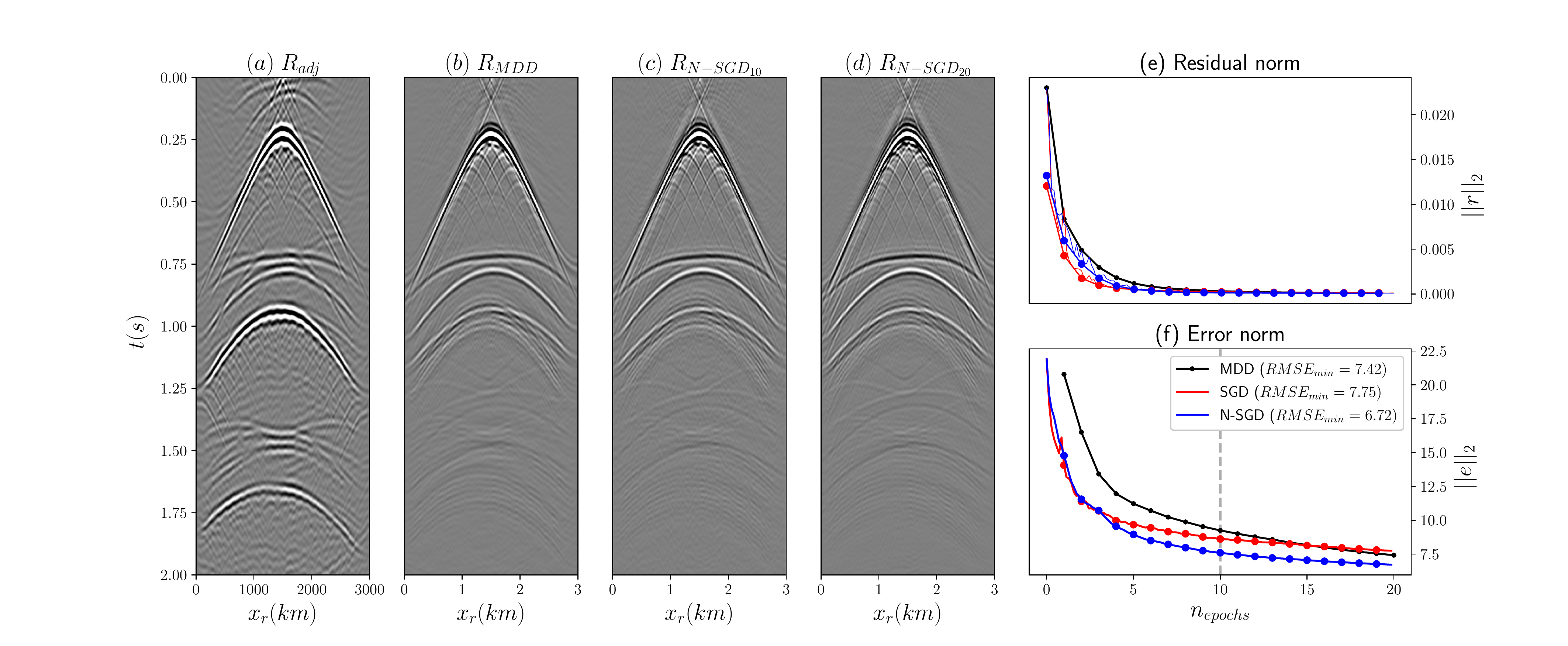}
  \caption{Reflection response estimates. a) Cross-correlation (i.e., adjoint), b) full-gradient MDD after 20 epochs, c) Stochastic MDD with N-SGD after 10 epochs, and d) Stochastic MDD with N-SGD after 20 epochs. e) Residual and f) error norms as function of epochs for the full-gradient MDD (black lines), MDD by means of SGD (red lines), and MDD by means of N-SGD (blue lines). In both plots, thick solid lines with dots display the values of the respective norms at the end of each epoch. Thin solid lines in panel e represent the norms at each iteration for the stochastic MDD algorithms (i.e., contain $N_s/N_{s, batch}$ elements per epoch). The norms of the stochastic MDD algorithm start from the first step and intermediate values for the iterations up the the end of the first epoch are also displayed, whilst those of the full-gradient MDD start from $n_{epochs}=1$.}
  \label{fig:obc_inversionnorms}
\end{figure}

A a first experiment, we perform redatuming for a single virtual source in the middle of the receiver array ($x_{VS}=1500m$). Given the high quality of the input wavefields and overall low level of noise in the data, additional constraints that couple the responses for multiple virtual sources are not required; apart from an improvement in terms of computational speed \cite{Ravasi2021}, similar results are in fact expected when inverting for multiple virtual sources simultaneously. The redatumed response obtained by applying the adjoint of the modelling operator to the up-going data in Fig. \ref{fig:obc_data}a is displayed in Fig. \ref{fig:obc_inversionnorms}a. Since the adjoint of the modelling operator can be equivalently interpreted as the cross-correlation between the up- and down-going wavefield, the resulting wavefield contains a mix of physical events with correct kinematic and spurious events due to the cross-talk between unrelated events in the up- and down-going wavefields \cite{wapenaar2011}. More specifically, when compared to the ideal reflection response in Fig. \ref{fig:obc_data}c, this wavefield contains strong spurious events at around $1s$ and $1.75s$ that we wish to remove when solving the inverse problem in equations \ref{eq:mddtinv} or \ref{eq:mddtfinite}. 

Three different inversions are carried out and shown in Fig. \ref{fig:obc_inversionnorms}, panels b to d. Fig. \ref{fig:obc_inversionnorms}b represents the benchmark reflection response obtained by means of full-gradient MDD after 10 iterations of LSQR. The following two panels display the reflection responses estimated by means of N-SGD for 10 and 20 epochs, respectively. In all cases, a batch size of $N_{s,batch}=32$ is used and the step-size selection is based on the semi-heuristic criterion described in Appendix B. Visually, the reflection responses retrieved by the various algorithms show a good agreement with the ideal reflection response in Fig. \ref{fig:obc_data}c. Importantly, the strong spurious events in Fig. \ref{fig:obc_inversionnorms}a have been successfully attenuated whilst the underlying weaker primary events are now visible. A closer inspection of the results, alongside a quantitative assessment of the reconstruction error of the various algorithms (Fig. \ref{fig:obc_inversionnorms}f), reveals a faster convergence of the stochastic MDD algorithms over the full-gradient MDD method. More specifically, if we compare the error norms after 10 epochs, we observe that both SGD and N-SGD produce a solution that is of superior quality to that of full-gradient MDD. Although the quality of the reflection response for full-gradient MDD further improves with the number of epochs and overtakes that of SGD at around 20 iterations, the result of N-SGD is still the one that produces the lowest RMSE. Such favourable behaviour can be explained with the fact that for each epoch of N-SGD we are performing $\lfloor N_s / N_{s, batch}\rfloor+1=7$ inexact gradient steps versus a single, although optimal, gradient step for LSRQ in full-gradient MDD. Finally, for this specific example, a similar behaviour is also observed for the norm of the residual, where the stochastic algorithms show a faster convergence to zero residual. Note that both iteration-wise norms (thin lines) and epoch-averaged norms (thick lines) are displayed in Fig. \ref{fig:obc_inversionnorms}e. Nevertheless, as we will see in later examples, the norm of the residual does not represent a reliable proxy metric for the error norm, which in real life examples cannot be directly evaluated. 

\begin{figure}
  \centering
  \includegraphics[width=0.9\textwidth]{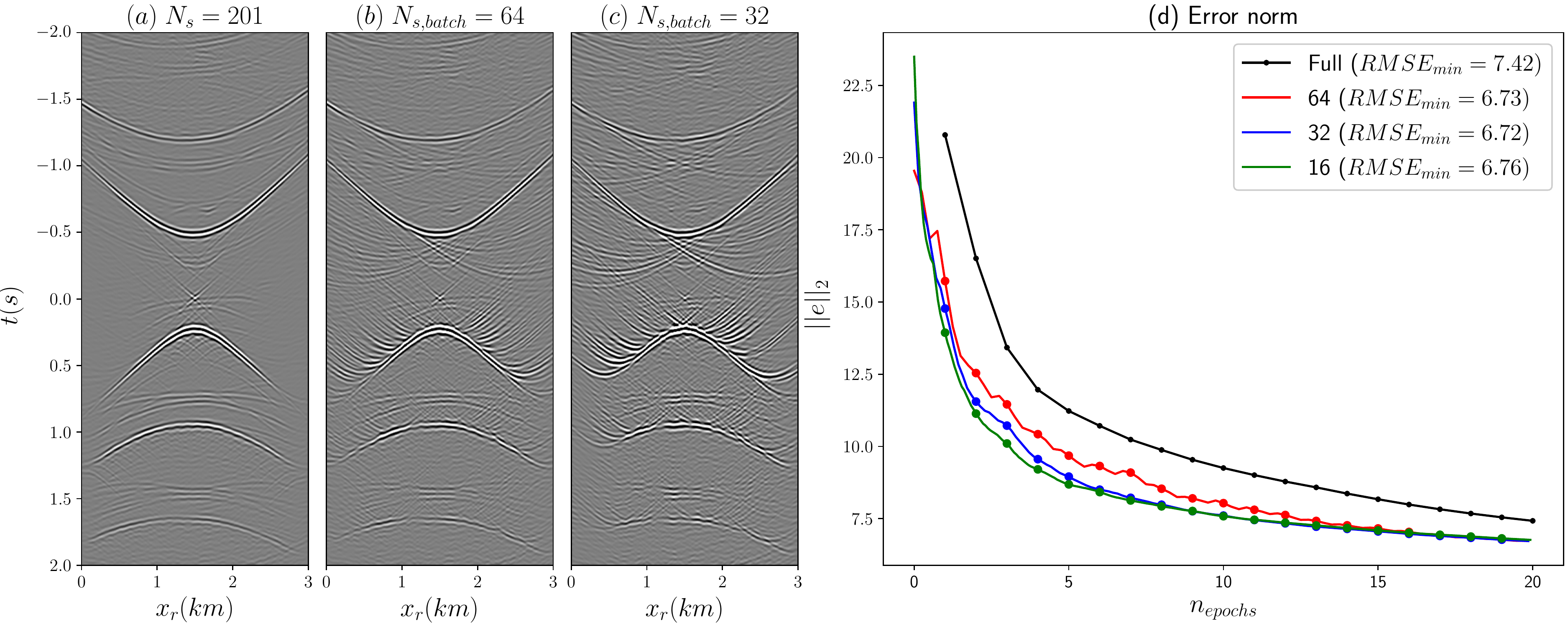}
  \caption{Gradients for a) full source array, b) batch of 64 randomly selected sources and c) batch of 32 randomly selected sources. d) Error norms as function of epochs for the various algorithms.}
  \label{fig:obc_gradientnorms}
\end{figure}

Finally, we want to assess the impact of batch size on the convergence of the stochastic MDD algorithms. This parameter plays a key role in that a too small batch size may produce highly corrupted gradients as well as introduce computational overhead due to the amount of forward and inverse FFTs additionally required in the evaluation of the gradients for the stochastic algorithms (see Appendix A for details). Nevertheless, a too large batch size may use redundant information in the generation of the gradients; moreover, although not directly shown in this paper, in 3D applications this may lead to out-of-core operations as the portion of the down-going wavefield used to compute the gradient may not fit in the available computer memory \cite{Ravasi2021}. Fig. \ref{fig:obc_gradientnorms} shows the initial gradient, i.e. $-\sum_{i_s \in S_{batch}} \textbf{P}_{t,i_s}^{+H} \textbf{p}_{t,i_s}^-$, for the full source array, and batches of 64 and 32 randomly selected sources. It is evident that the main features of the full-gradient are preserved in the approximate gradients, nevertheless some spurious hyperbolic events with opposite curvature arise due the poor sampling of stationary points. The behaviour of the error norms (Fig. \ref{fig:obc_gradientnorms}) does however indicate that the smaller number of sources used to compute each gradient the faster the overall convergence, leading to a trade-off between computational efficiency in the modelling operator and convergence speed in the choice of the batch size.

\subsection{Synthetic Subsalt redatuming}
\begin{figure}
  \centering
  \includegraphics[width=0.7\textwidth]{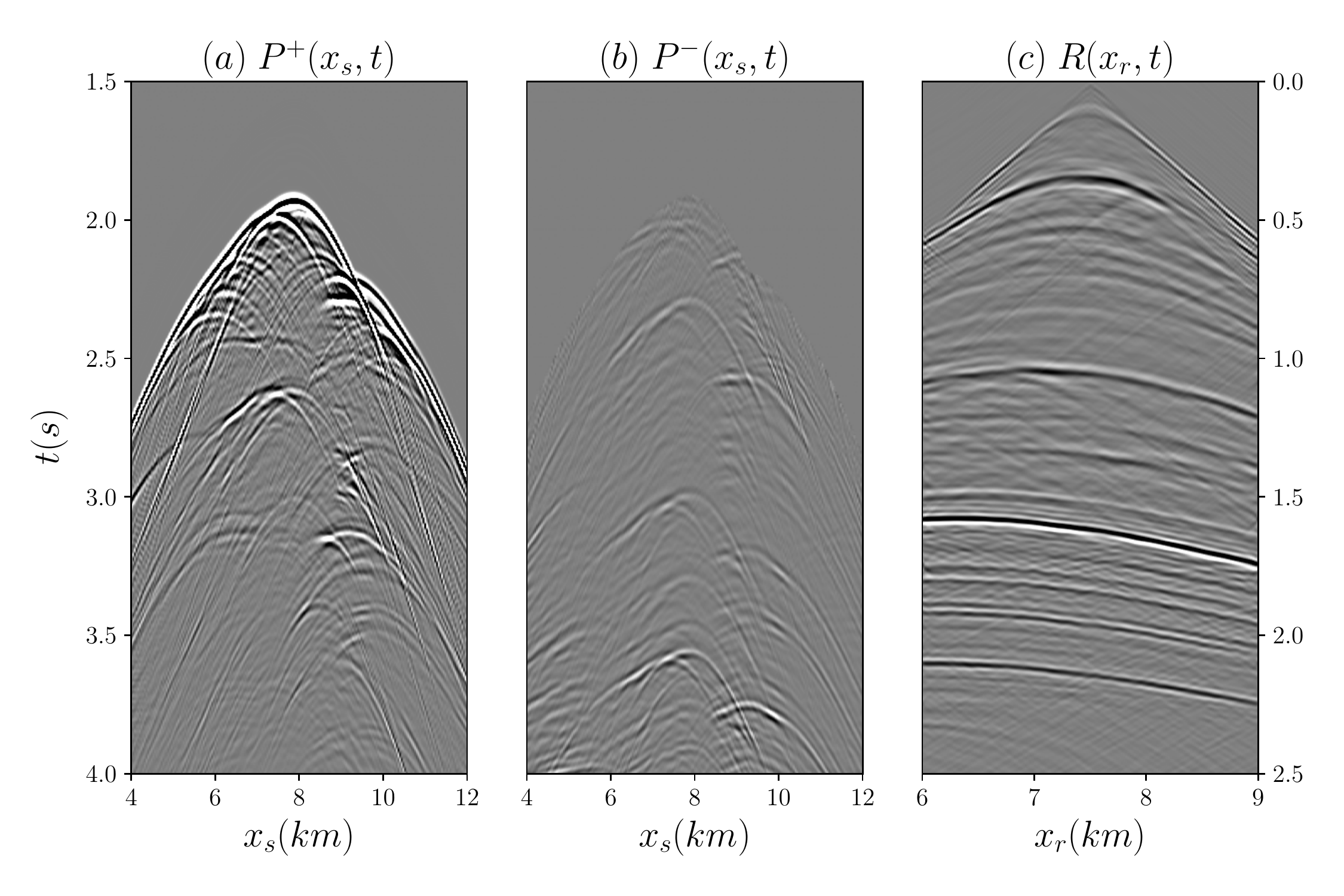}
  \caption{a) Down- and b) up-going common-receiver gathers for a receiver in the middle of the subsurface array. c) Ideal reflection response for a virtual source in the middle of the subsurface array.}
  \label{fig:salt_data}
\end{figure}

In this second example, we consider a scenario of source-side, target-oriented redatuming below a salt body. Here the input up- and down-going wavefields are not simply obtained by means of wavefield separation, rather they are the product of a step of receiver-side redatuming of surface data. More specifically, the up- and down going separated surface data are used as input to the SRM redatuming scheme \cite{vargas2021a}; a highly ill-posed inverse problem that despite its high degree of accuracy when compared to other redatuming methods (e.g., \cite{berryhill1984, Wapenaar2014}), inevitably introduces errors in the form of both coherent and incoherent noise in the redatumed responses. Fig. \ref{fig:salt_data} displays the down- and up-going wavefields for a receiver in the middle of the subsurface array, alongside the ideal reflection response deprived of any overburden effect (i.e., modelled in a truncated medium starting from a depth of $z=4500m$). 


\begin{figure*}[ht]
  \centering
  \includegraphics[width=.9\textwidth]{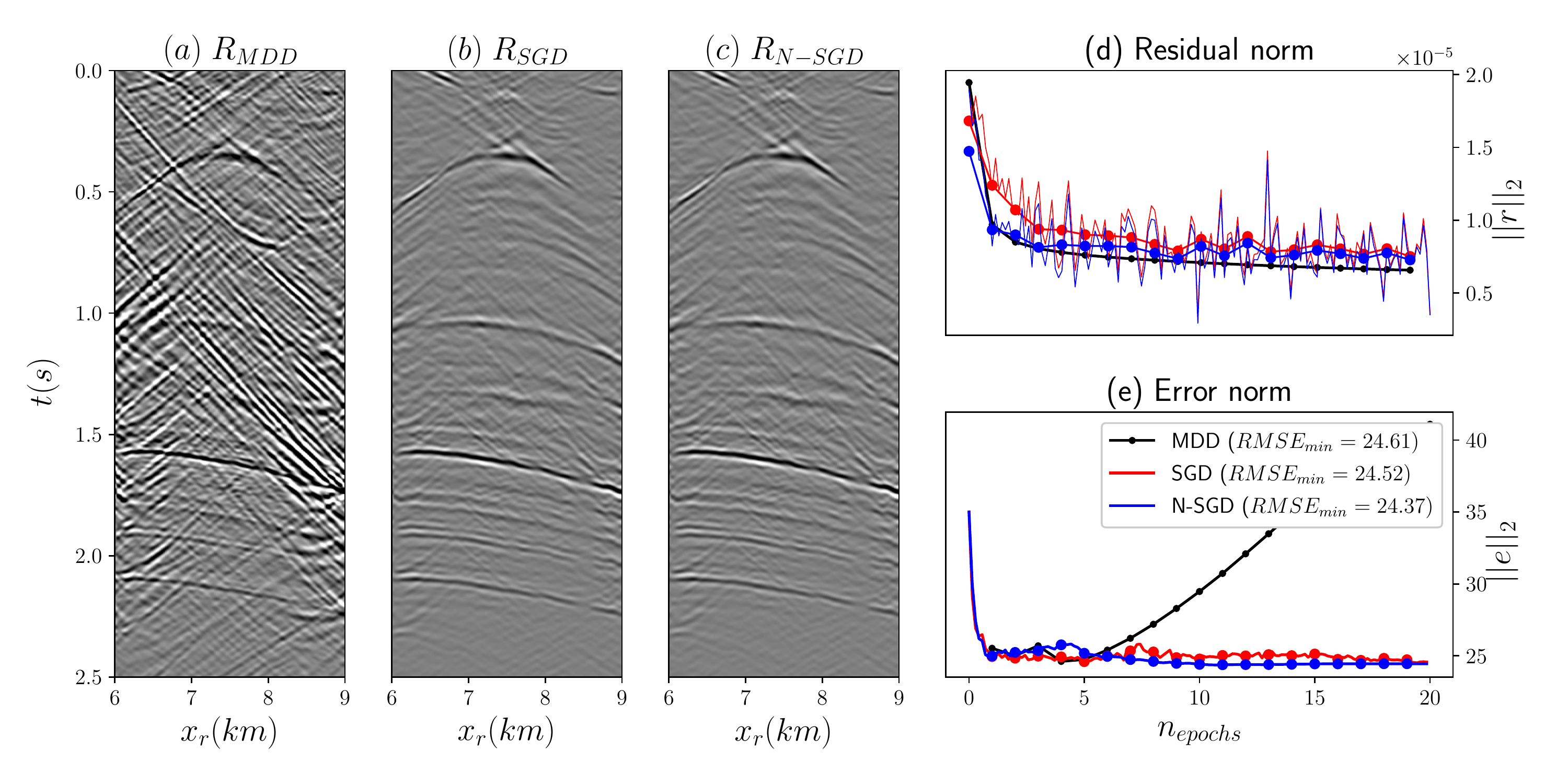}  
  \caption{Reflection response estimates after 20 epochs for single virtual source inversion. a) Full-gradient MDD, b) Stochastic MDD with SGD, c) Stochastic MDD with N-SGD d) Residual and e) error norms as function of epochs. Keys as in Fig. \ref{fig:obc_inversionnorms}.}
  \label{fig:salt_singleinversion}
\end{figure*}

\begin{figure}
  \centering
  \includegraphics[width=.9\textwidth]{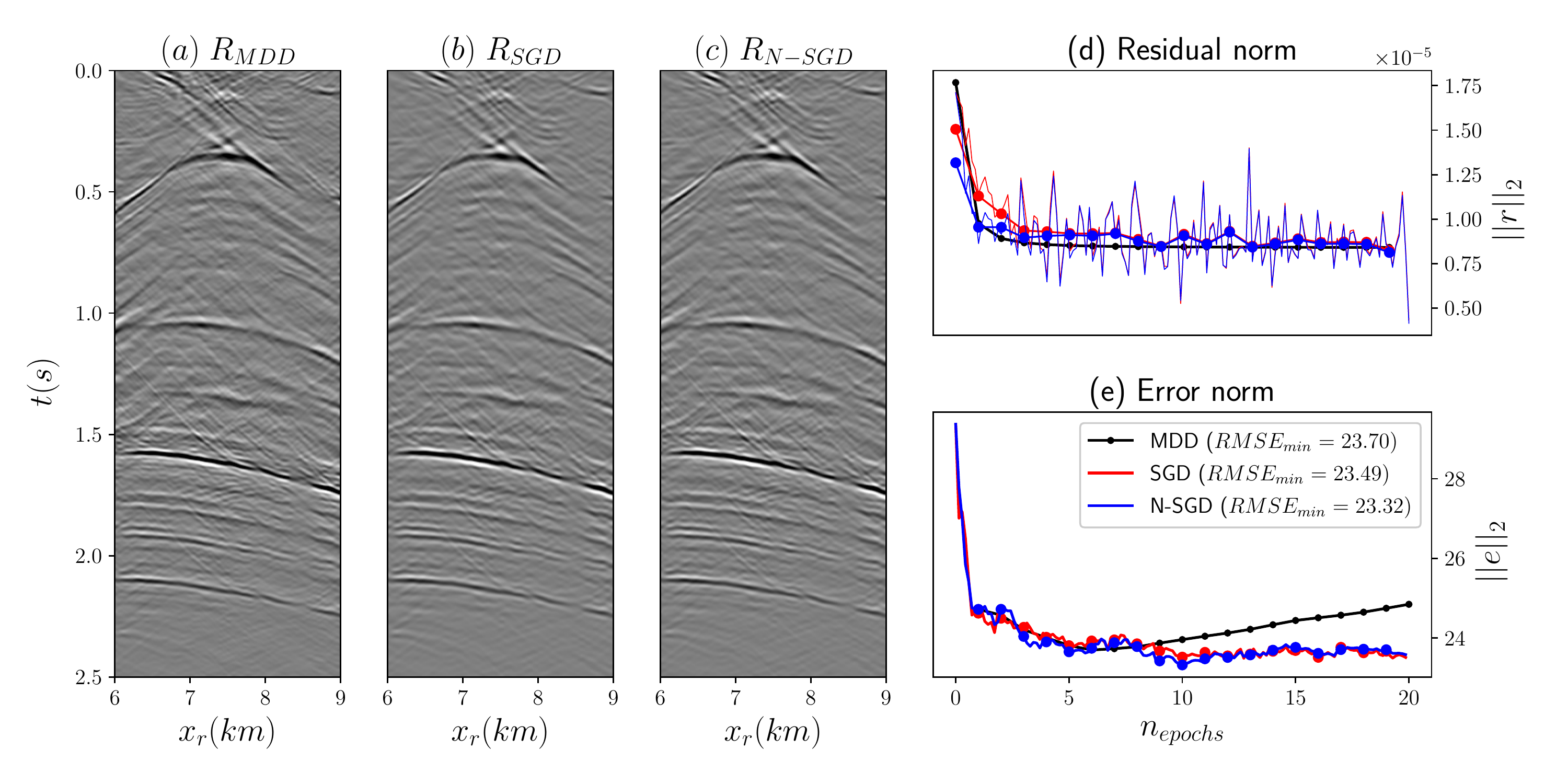}  
  \caption{Reflection response estimates after 20 epochs for single virtual source inversion. a) Cross-correlation , b) Single-virtual-source full-gradient MDD, c) Multi-virtual-source full-gradient MDD, d) Single-virtual-source stochastic MDD with N-SGD, and e) Multi-virtual-source stochastic MDD with N-SGD. Keys as in Fig. \ref{fig:obc_inversionnorms}.}
  \label{fig:salt_multiinversion}
\end{figure}

\begin{figure}
  \centering
  \includegraphics[width=0.6\textwidth]{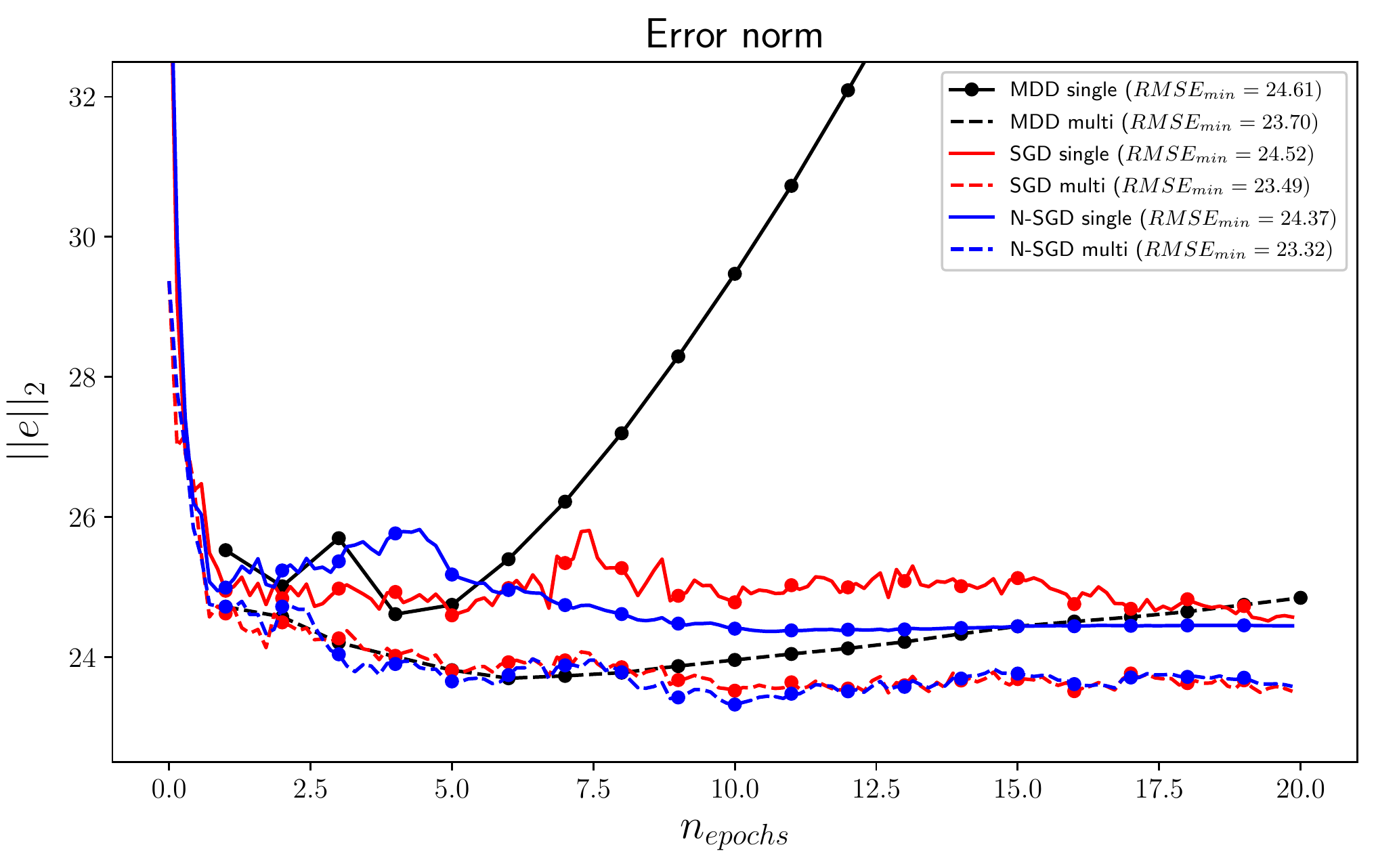}
  \caption{Close-up of error norms for the different algorithms. Solid and dashed lines are used for single- and multi-virtual strategies, respectively.}
  \label{fig:salt_normszoom}
\end{figure}

As recently discussed in \cite{vargas2021b}, this dataset presents a perfect playground for the development of robust MDD algorithms; we now aim to assess the capabilities of stochastic MDD in the presence of noise in both the data and modelling operator, when inverting for either one or multiple virtual sources. The former approach presents the clear advantage of being less computationally demanding and easier to parallelize over multiple virtual sources; the latter provides the opportunity to introduce a reciprocity physical constraint, however it comes with a number of additional computational challenges due to the size of the model and data vectors to be used in the inversion process. The redatumed wavefields obtained using a single virtual source in the middle of the array are displayed in Fig. \ref{fig:salt_singleinversion}. The response retrieved by means of full-gradient MDD (Fig. \ref{fig:salt_singleinversion}a) is clearly contaminated by strong dipping noise, which arises due to the fact that both the data and modelling operator are inexact. By computing the norm of the model error through iterations, we can see how the black line in Fig. \ref{fig:salt_singleinversion}e decreases in the first few iterations, whilst it starts to grow as artefacts become more prominent in later iterations. This behaviour, usually referred to as semi-convergence \cite{hansen2010}, has been also observed in \cite{luiken2020}. Although this result highlights the importance of early stopping in full-gradient MDD, in practical applications the error norm cannot be computed; no indication of data overfitting can be observed by looking at a proxy norm measurement such as the residual norm (black line in Fig. \ref{fig:salt_singleinversion}d), making it difficult to objectively and automatically decide the optimal number of iterations of MDD. On the other hand, the reflection responses produced by means of stochastic MDD, with either SGD or N-SGD, are much cleaner and ultimately closer to the true response (Figs. \ref{fig:salt_singleinversion}c and d). Assuming availability of computational resources, which for 2D examples is within reach of today's computers, the full-gradient and stochastic MDD algorithms are now used to simultaneously invert for the entire set of virtual sources (Fig. \ref{fig:salt_multiinversion}). This allows for the introduction of a reciprocity preconditioner, which is very beneficial to this kind of highly ill-posed MDD problems. Clearly, the reflection response retrieved by full-gradient MDD is of a much higher quality compared to the previous result obtained by using a single virtual source. The reflection responses from the stochastic MDD methods are also very clean, and of slightly higher quality when compared to their full-gradient counterpart. Fig.~\ref{fig:salt_normszoom} shows a close up of the error norms for the six different inversion results shown in Figs. \ref{fig:salt_singleinversion} and \ref{fig:salt_multiinversion}. It can be observed how, in general, the introduction of a reciprocity preconditioner improves the solution of both full-gradient and stochastic MDD. However, whilst this seems to be of vital importance for the former method, the improvement is fairly minor in the latter, making it possible to work with a single or small group of virtual sources at the same time without seriously compromising the quality of the reconstruction. As already discussed above, this finding is of great importance for the application of MDD to large-scale, three dimensional datasets where solving for the entire set of virtual sources at the same time may be beyond reach of our current compute capabilities.

\subsection{Volve OBC Up/Down Deconvolution}
In this final example, we consider a 2D line of the open-source Volve dataset. Volve is an oil field located in the central part of the North Sea, five kilometres north of the Sleipner Øst field. The field was shut down in 2016, with the facility removed in 2018, and the historical subsurface and production data were made available by Equinor and partners in June 2018. 

\begin{figure}
  \centering
  \includegraphics[width=0.7\textwidth]{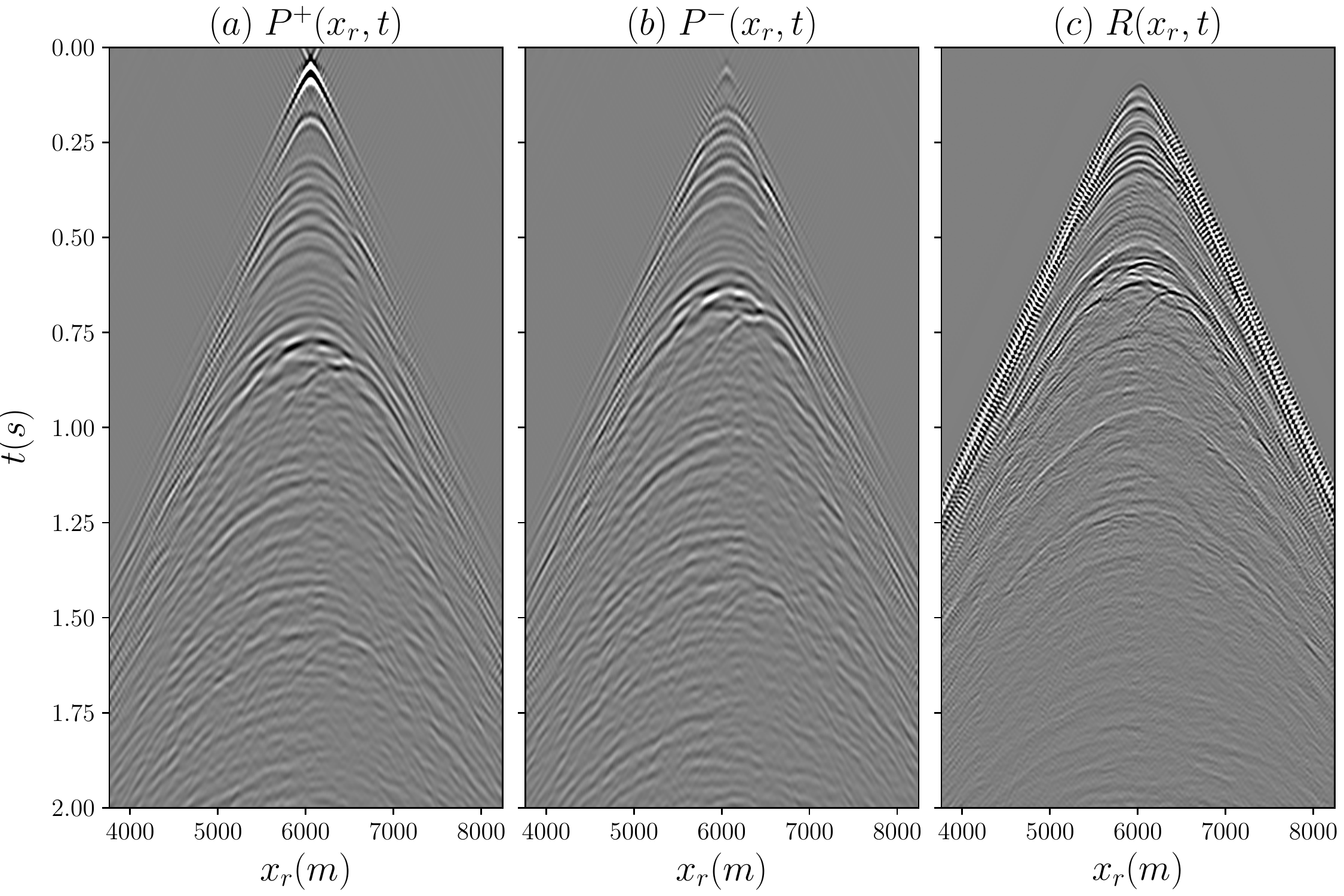}
  \caption{Common shot gather in the middle of the source array for the a) down-going wavefield, b) up-going wavefield, and c) target reflection response without free-surface effects.}
  \label{fig:volvesynth_data}
\end{figure}

In order to be able to assess the performance of stochastic MDD against full-gradient MDD, we begin by creating a synthetic, Volve-like dataset. The dataset creation is composed of 3 main steps:

\begin{itemize}
    \item The migration velocity model in Fig. \ref{fig:models}d (\textit{ST10010ZC11-MIG-VEL.MIG\_VEL.VELOCITY.3D.JS-017527.segy}) is first converted into an equivalent acoustic impedance background model with the help of available well logs by means of linear regression;
    \item The 3D post-stack seismic dataset \textit{ST10010ZC11\_PZ\_PSDM\_KIRCH\_FULL\_D.MIG\_FIN} \textit{.POST\_STACK.3D.JS-017536.segy} is used alongside the background acoustic impedance model as input to a step of post-stack inversion (see \cite{Ravasi2021b} for more details); the resulting detailed acoustic impedance model is converted back to an equivalent velocity model using the inverse of the velocity-acoustic impedance relation at well locations;
    \item Pressure and particle velocity seismic data are modelled using a Ricker wavelet with central frequency $f_c=20Hz$ and a free-surface at the top of the model. A line of 110 sources at depth of $z_s=6m$ with spatial sampling $dx_s=50m$ is used alongside a line of 180 receivers at varying depth ranging from $z_{r,min}=86m$ to $z_{r,max}=99m$ and spatial sampling $dx_r=25m$.
    Similarly, the reference reflection response is modelled by filling the water layer with the velocity of the seafloor and deactivating the free-surface. In this case, both sources and receivers are placed along the seafloor (Fig. \ref{fig:volvesynth_data}c).
\end{itemize}

\begin{figure}
  \centering
  \includegraphics[width=0.99\textwidth]{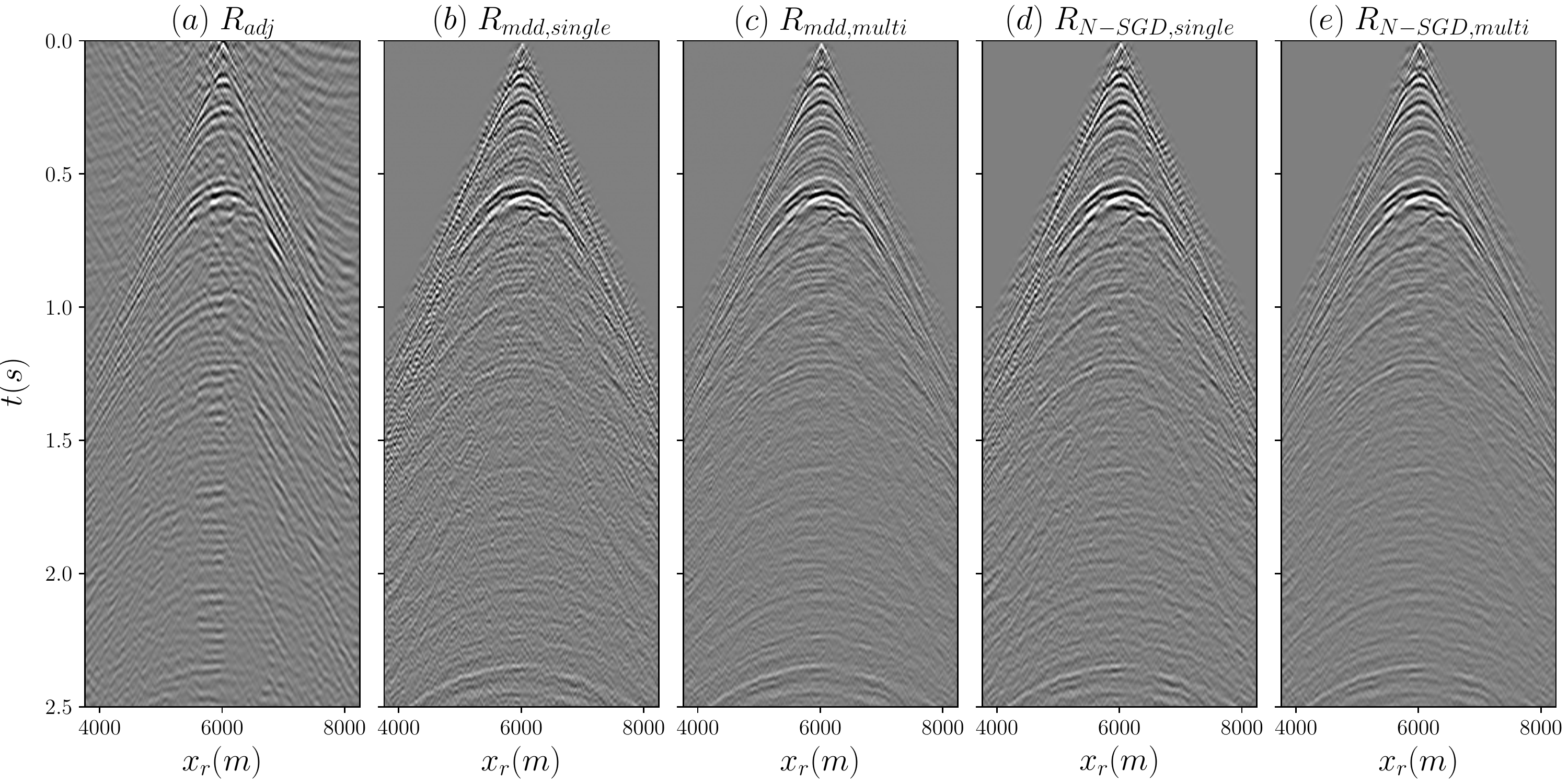}
  \caption{Reflection response estimates after 30 epochs. a) Full-gradient MDD, b) Stochastic MDD with SGD, c) Stochastic MDD with N-SGD d) Residual and e) error norms as function of epochs. Keys as in Fig. \ref{fig:obc_inversionnorms}.}
  \label{fig:volvesynth_inv}
\end{figure}

\begin{figure}
  \centering
  \includegraphics[width=0.6\textwidth]{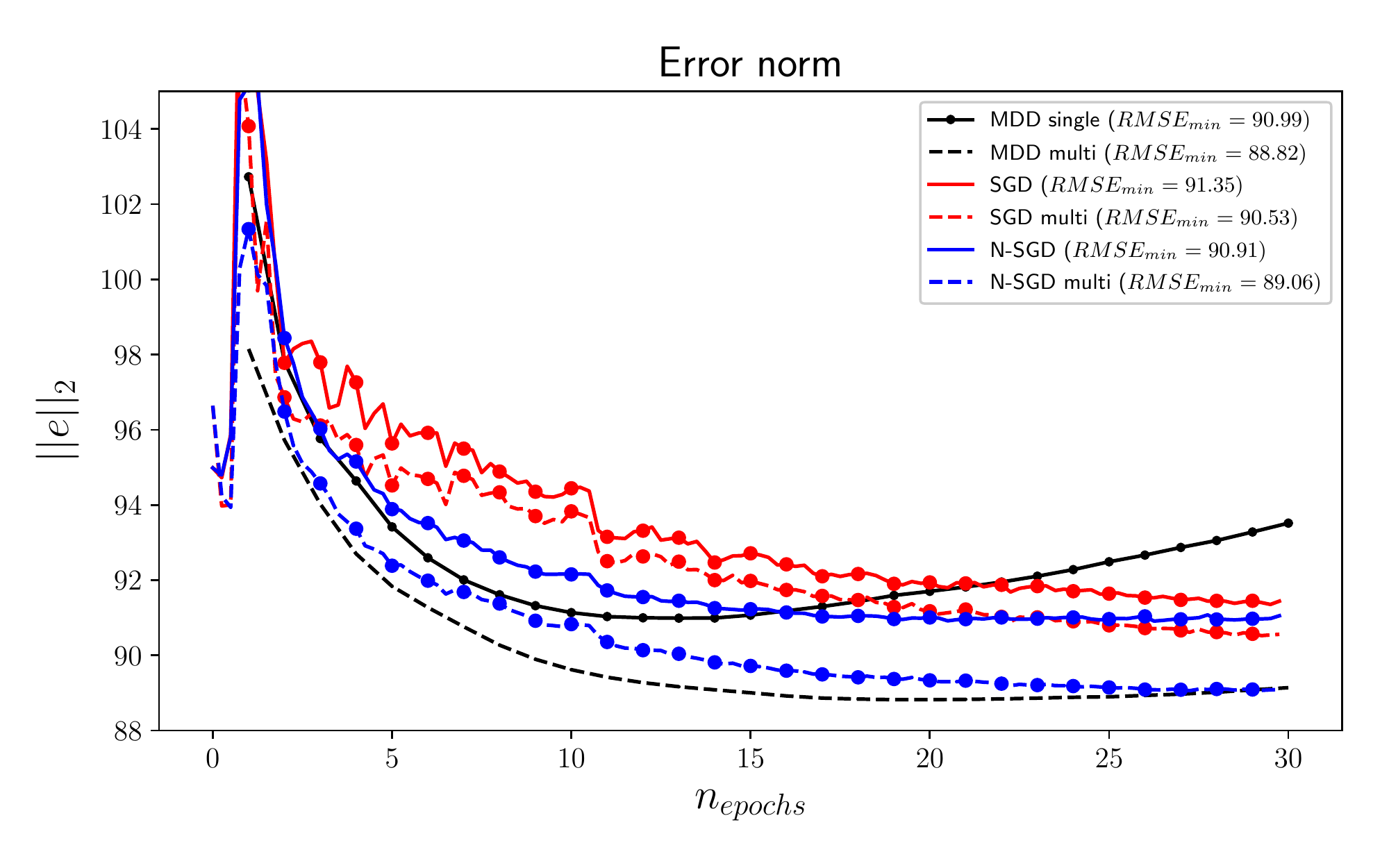}
  \caption{Close-up of error norms for the different algorithms. Solid and dashed lines are used for single- and multi-virtual strategies, respectively.}
  \label{fig:volvesynth_norms}
\end{figure}

Similar to previous experiments, the up- and down-going pressure wavefield are created by means of wavefield separation (Figs. \ref{fig:volvesynth_data}a and b) and used as input to the various MDD algorithms. It must be noted here that, whilst wavefields are of much greater complexity than those in the first example, the amount of noise introduced by the wavefield separation process is negligible when compared to that introduced by a previous step of redatuming, like in the subsalt example. In other words, we expect both full-gradient and stochastic MDD to perform well in this scenario; our main interest lies in understanding the stability and convergence of the various MDD processes as a proxy to what we will observe in the field data experiments (without a reference solution to compare our results onto). Fig. \ref{fig:volvesynth_inv} displays the reconstructed reflection responses for a virtual source in the middle of the receiver array using the adjoint of the modelling operator (Fig. \ref{fig:volvesynth_inv}a), single- and multi-virtual source full-gradient MDD (Figs. \ref{fig:volvesynth_inv}b and c), and single- and multi-virtual source stochastic MDD (Figs. \ref{fig:volvesynth_inv}d and f). All inversions are carried out with a causality preconditioner (i.e., a space-time mask that mutes amplitudes above the direct arrival in the reconstructed reflection responses); moreover, an additional reciprocity preconditioner is used for multi-virtual source inversions. Given the presence of a fairly shallow seabed, cross-talk between unrelated primaries and free-surface multiples is visible in the reflection response constructed by means of cross-correlation. Such spurious events are clearly suppressed in the MDD results independently on the choice of the algorithm. Nevertheless, given the complexity of the wavefield and therefore the ill-posed nature of the problem, significant incoherent noise is introduced in the solution obtained from single-virtual-source full-gradient MDD (Fig. \ref{fig:volvesynth_inv}b). Such noise is visibly reduced when introducing a reciprocity preconditioner as part of the multi-virtual-source full-gradient MDD (Fig. \ref{fig:volvesynth_inv}c). The reflection responses obtained by means of stochastic MDD corroborate our previous findings that working with batches of sources seems to act as a natural regularizer to the final solution. Once again, working with multiple virtual sources and introducing a reciprocity preconditioner further improves the quality of the solution; however, the improvement in this case is much less evident. Similar observations can be drawn from Fig. \ref{fig:volvesynth_norms}, where we observe the usual semi-convergence behaviour for the full-gradient MDD solutions and a more stable convergence for the stochastic MDD cases. Nevertheless, in this specific example, the convergence of full-gradient MDD, especially when using multiple virtual sources in combination with a reciprocity preconditioner, seem to slightly outperform that of the stochastic MDD algorithms.

\begin{figure*}[htb]
  \centering
  \includegraphics[width=0.9\textwidth]{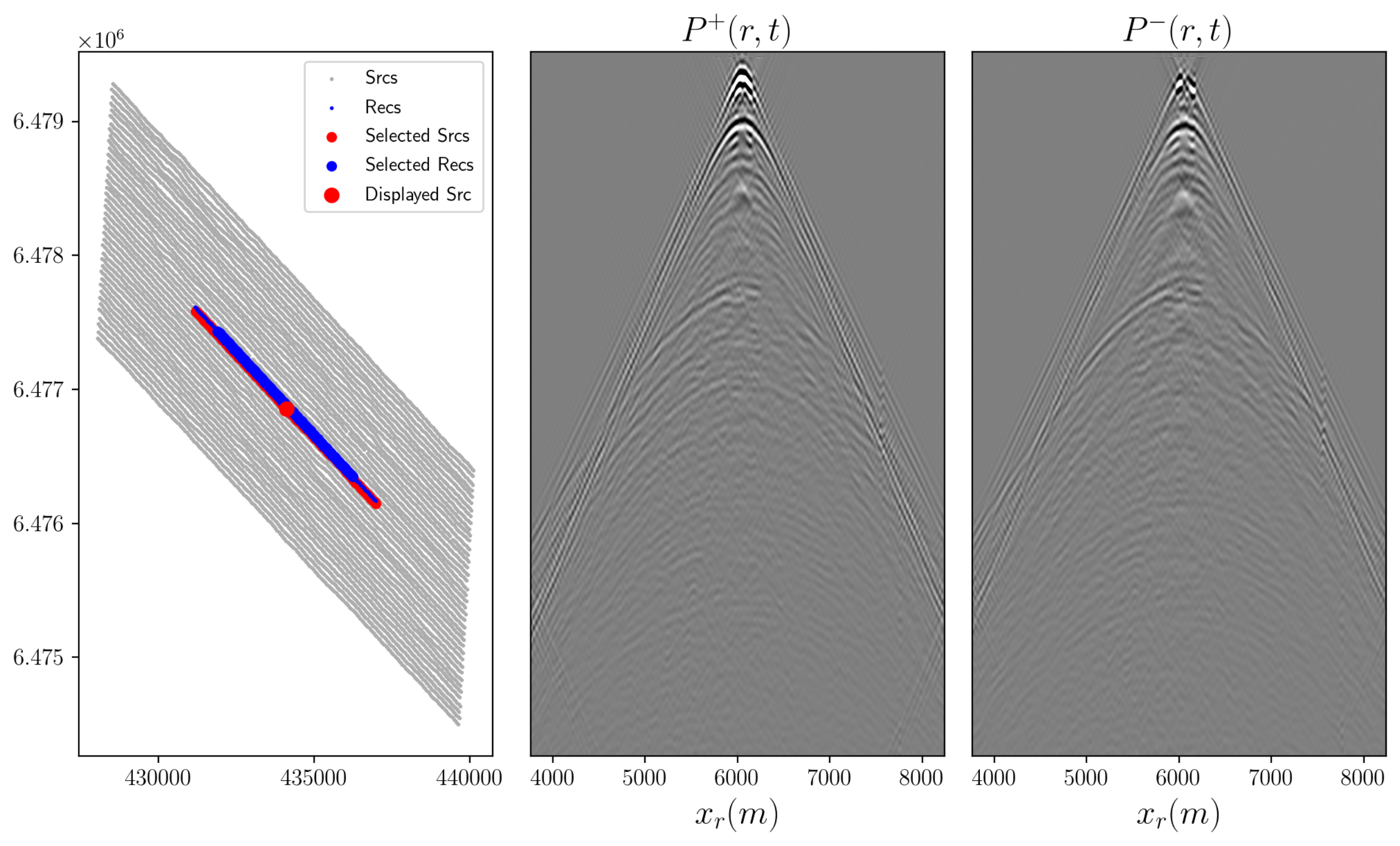}
  \caption{a) Acquisition geometry with selected 2D line of sources (red dots) and receivers (blue dots). b) Down-going and c) Up-going pressure data for a single source in the middle of the 2D line (large red dot in panel a).}
  \label{fig:volve_data}
\end{figure*}

\begin{figure*}[htb]
  \centering
  \includegraphics[width=0.9\textwidth]{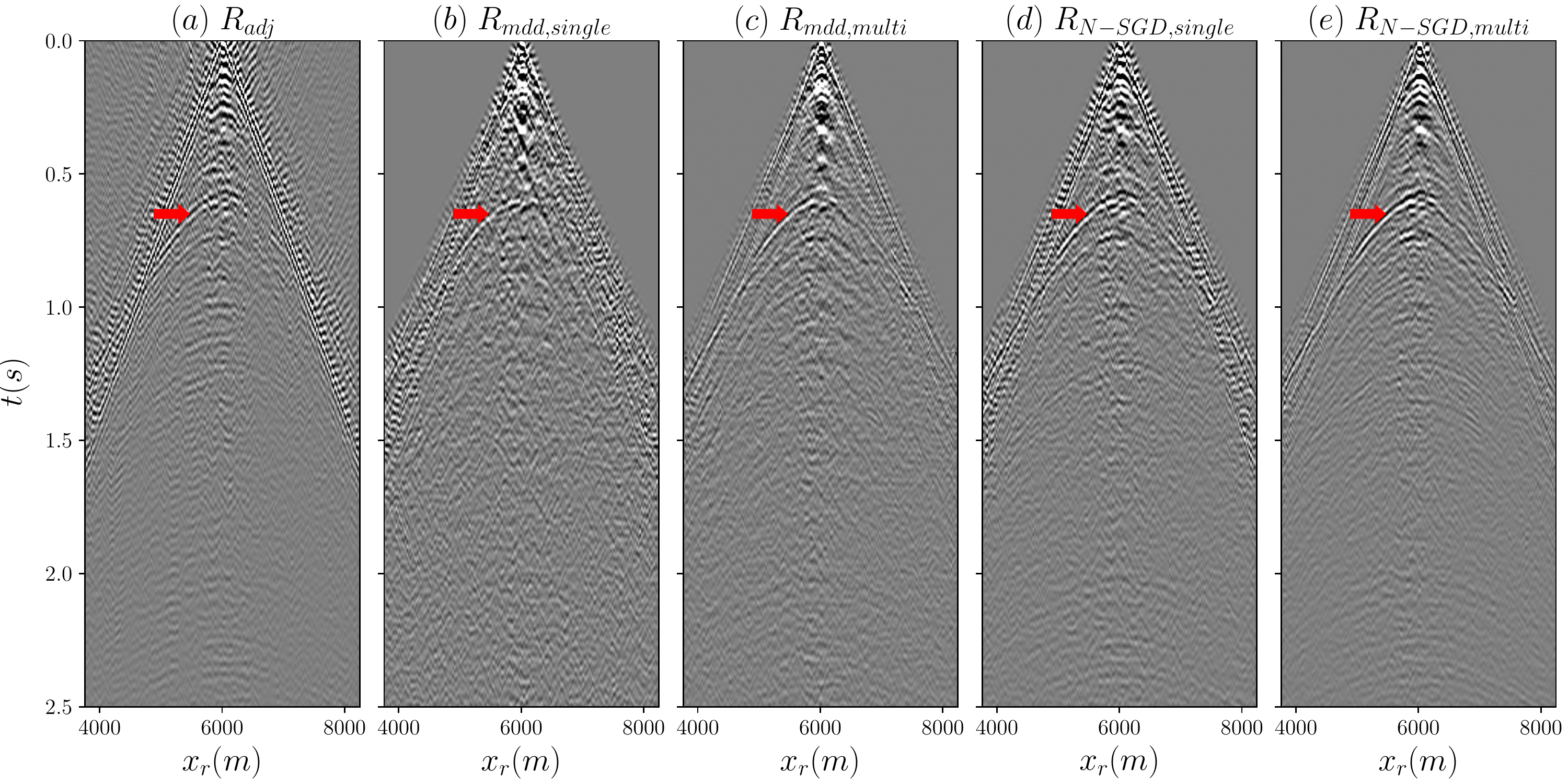}
  \caption{Reflection responses estimated by a) cross-correlation b) single-virtual-source full-gradient MDD, c) multi-virtual-source full-gradient MDD, d) single-virtual-source N-SGD MDD, and e) multi-virtual-source N-SGD MDD. All MDD estimates are produced using X epochs.}
  \label{fig:volve_mdd}
\end{figure*}

\begin{figure*}[htb]
  \centering
  \includegraphics[width=0.9\textwidth]{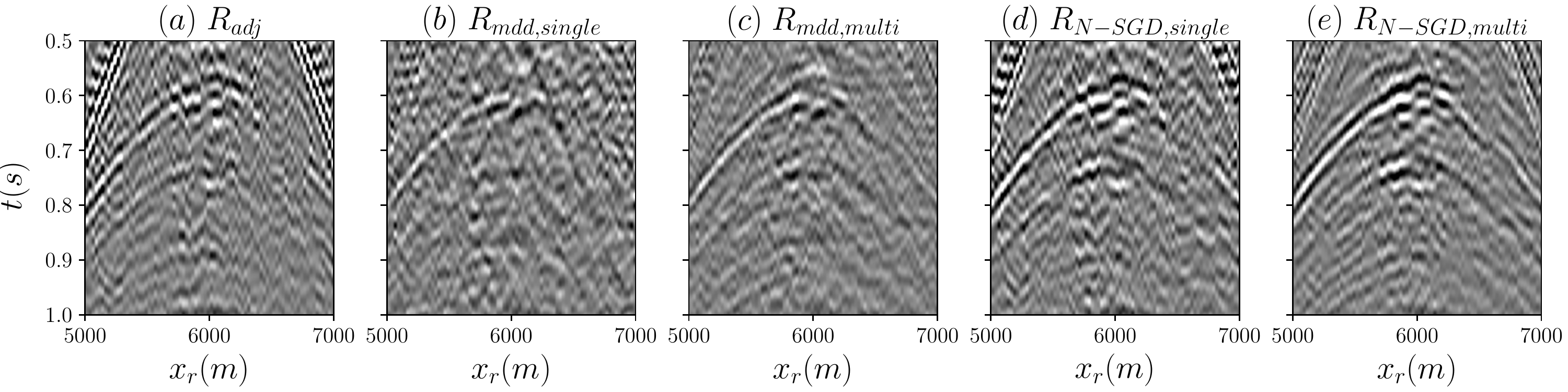}
  \caption{Close-up of the reflection responses ordered in the same fashion of Fig. \ref{fig:volve_mdd}.}
  \label{fig:volve_mddzoom}
\end{figure*}

\begin{figure*}[htb]
  \centering
  \includegraphics[width=0.99\textwidth]{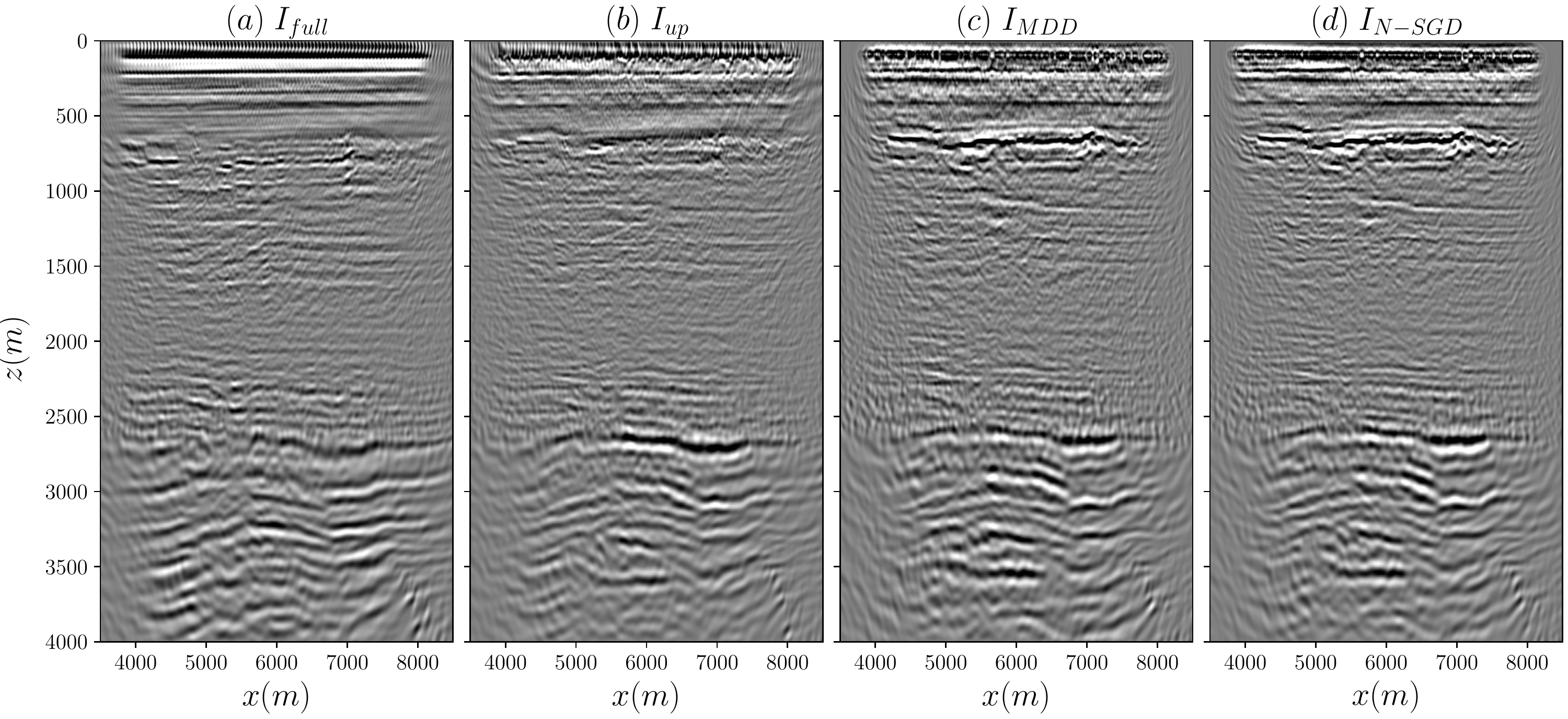}
  \caption{Images obtained using a) full pressure data, b) up-going pressure wavefield, c) reflection response from multi-virtual-source full-gradient MDD, and d) reflection response frommulti-virtual-source N-SGD MDD.}
  \label{fig:volve_images}
\end{figure*}

Moving onto the field dataset, a 2D line of sources and receivers is selected from the \textit{ST10010\_1150780\_40203.sgy} file that contains a portion of the 3D OBC dataset acquired by Statoil in 2010 (Fig. \ref{fig:volve_data}a). The pressure and vertical particle velocity components are pre-processed using a similar flow to that described in \cite{ravasi2015} and \cite{Ravasi2016}, and the resulting up- and down-going wavefields for a source in the middle of the 2D acquisition geometry are displayed in Figs. \ref{fig:volve_data}b and c. Full-gradient and stochastic MDD are applied to the separated wavefield for 40 epochs. Similar to the synthetic example, causality and reciprocity preconditioners are used to aid the solution of both MDD algorithms. Fig. \ref{fig:volve_mdd} displays the estimated reflection responses by means of cross-correlation (Fig. \ref{fig:volve_mdd}a) and various MDD algorithm (Figs. \ref{fig:volve_mdd}b-e). Note that whilst the different MDD algorithms are all able to some extent to remove the cross-talk artefacts visible in the cross-correlation gather, the stochastic MDD algorithms produce cleaner wavefields (see, for example, the event indicated by red arrows). The quality of the different reconstructions can also be appreciated in a close-up taken from around $0.8s$ (Fig. \ref{fig:volve_mddzoom}). Finally, to better assess the impact of MDD and the difference between the full-gradient and stochastic algorithms, the full and up-going pressure data as well as the estimated reflection responses from the multi-virtual-source MDD algorithms are all imaged by means of Reverse-Time Migration (Fig. \ref{fig:volve_images}). We can clearly observe that both the receiver ghost and higher-order free-surface multiples are successfully suppressed in the two MDD images. Moreover, the image created from the reflection response estimated by means of stochastic MDD is visibly cleaner especially in the shallow part of the subsurface.

\section*{Discussion}

Stochastic optimization algorithms have been popularized in geophysics by \cite{vanLeeuwen2011} and \cite{vanHerwaarden2020} in the context of partial differential equation-based (PDE-based) nonlinear inverse problems, i.e. full waveform inversion (FWI). In such a scenario, the cost of computing a gradient is driven by the number of PDE solves required to evaluate a chosen objective function; the benefit of using inexact gradients is represented by the fact that nearby sources provide highly redundant information in the construction of the gradient itself. Moreover, as the problem is inherently non-convex, there is no guarantee to reach the global minimum irrespective of the fact that full or partial gradients are used.

In our work, we are instead dealing with a linear inverse problem, which ideally has a unique minimizer; this may cast some doubts onto the benefit of using stochastic solvers. It is however important to remember that the forward operator of the MDD problem is highly ill-posed  -- i.e., the kernel matrix is rank deficient and has a large nullspace as shown, for example, in Fig. \ref{fig:steepestobc}a. Therefore, a variety of solutions exist that match the data equally well. Moreover, similar to FWI, the information carried by each source to the gradient can be redundant. Our numerical experiments highlight two main benefits in the use of stochastic solvers for MDD: robustness to noise in the data and operator, and faster convergence at times. The latter benefit is clearly highlighted in the first synthetic example where high-fidelity up- and down-going wavefields are used as input to the MDD problem. As negligible noise affects such measurements (even after wavefield separation), both the full and stochastic solvers converge to an accurate solution; nevertheless, the convergence of the N-SGD solver is faster by a factor of nearly two in comparison to full-gradient MDD.

A major difference between full-gradient and stochastic gradient-based algorithms is that the former optimizes a constant functional whilst the cost function changes from step to step in the latter. Although the use of large batch sizes can provide an unbiased estimate of the gradient, no analytical, exact formula exists for the step size at each iteration. This is obviously not the case in full-gradient iterative solvers, like CGLS \cite{shewchuk1994} or LSQR. The choice of the step size becomes therefore a very important hyperparameter to the success of stochastic MDD. The problem here is two fold: first, we wish to find the best (or a satisfactory) step size that provides fast and, at the same time, stable convergence. Second, we would like to be able to identify this scaling factor upfront without the need for trial-and-error that would inevitably narrow the computational gap between the full and stochastic MDD. The semi-heuristic procedure based on the Landweber iteration (see Appendix B) is shown in the numerical examples to be effective in this context.

\section*{Conclusions}
We have presented a novel formulation of Multi-Dimensional Deconvolution that recast the associated inverse problem into an optimization of a finite-sum functional. By doing so, its solution is obtained by means of stochastic gradient descent algorithms, allowing gradients at each step to be computed using a (small) random subset of sources. Synthetic and field data examples validate the effectiveness of the proposed methodology, and its superiority over full-gradient MDD, both in terms of convergence speed and quality of the estimated local reflection response. As a by-product of an overall more stable convergence, stochastic MDD is also shown to rely less heavily on inter-virtual-source constraints (i.e., reciprocity preconditioning), ultimately easing its extension to large-scale, three dimensional datasets.

\appendix

\section{Computational cost of Stochastic MDD}
In this Appendix, we compare the computational cost of full-gradient and stochastic MDD. It must be noted that, unlike FWI based on random shot selection, our modeling operator is only partially separable among shots and therefore some additional overhead arises when splitting it into group of sources.

To being with, let's revisit the computational cost of the Multi-Dimensional Convolution operator in equation \ref{eq:mddintegral}. This operator is composed of three consecutive steps, namely forward Fourier transform (along the time axis), batched matrix-matrix multiplication (MMM) -- which reduces to matrix-vector multiplication (MVM) for the single-virtual-source case -- and inverse Fourier transform (along the time axis), whose computational costs are:
\begin{equation}
\label{eq:mdccost}
\begin{split}
&FFT: \mathcal{O}(N_t \cdot log_2(N_t)\cdot N_r \cdot N_{vs}) \\
&Batched\;MMM: \mathcal{O}( N_t \cdot N_s \cdot N_r \cdot N_{vs}) \\
&IFFT: \mathcal{O}(N_t \cdot log_2(N_t)\cdot N_s \cdot N_{vs}) \\
\end{split}
\end{equation}
where we assume here for simplicity that the number of samples in the Fourier domain equals that of the time domain.

In the case of full-gradient MDD, the cost of a single epoch is therefore equal to twice the cost of the MDC operator, as a forward and adjoint pass is required at each step of the solver:
\begin{equation}
\label{eq:mddcost}
\begin{split}
Full\; MDD\;(1\;epoch):\\
2\cdot [\mathcal{O}(N_t \cdot log_2(N_t)\cdot N_r \cdot N_{vs}) +\\ \mathcal{O}( N_t \cdot N_s \cdot N_r \cdot N_{vs}) +\\
\mathcal{O}(N_t \cdot log_2(N_t)\cdot N_s \cdot N_{vs})]
\end{split}
\end{equation}

On the other hand, the cost of an epoch of stochastic MDD equals $2N_b$ the cost of a MDC operation with $N_{s,batch}$ sources, where $N_b$ is the overall number of batches of sources. In this case, $N_{s,batch}$ is also used in spite of $N_s$ in the definition of the cost of the Batched MMM and IFFT steps of MDC operator. The cost of the FFT step is however the same as for the full MDC operator as it is independent on the source axis. An explicit expression for the cost of an epoch of stochastic MDD is:
\begin{equation}
\label{eq:stochmddcost}
\begin{split}
Stochastic\; MDD\;(1\;epoch):\\
2\cdot N_b \cdot [\mathcal{O}(N_t \cdot log_2(N_t)\cdot N_r \cdot N_{vs}) +\\
\mathcal{O}( N_t \cdot N_{s,batch} \cdot N_r \cdot N_{vs}) +\\
\mathcal{O}(N_t \cdot log_2(N_t)\cdot N_{s,batch} \cdot N_{vs})]= \\
(N_b-1) \cdot \mathcal{O}(N_t \cdot log_2(N_t)\cdot N_r \cdot N_{vs}) \\
+ Full\; MDD\;(1\;epoch)
\end{split}\end{equation}
This result shows that an additional $(N_b-1)$ forward Fourier transforms are performed at each epoch of the stochastic MDD algorithm compared to its full-gradient counterpart. However, when $N_s \gg log_2(N_t)$, the second step of the MDC operator dominates in terms of computational cost. This is a common scenario as the number of time samples of a seismic recording is usually in the order of $10^3$ ($log_2(N_t) \alpha 10$) whilst the number of sources can be around $10^2-10^3$ for 2D acquisition systems, and $10^3-10^5$ for 3D acquisition systems.

\section{Selecting step-size in stochastic gradient descent}
A key factor in the success of stochastic MDD is represented by the choice of the step-size. A too small step-size would inevitably lead to slow convergence, whilst a divergent solution would be produced when choosing a too large step-size. Obviously this parameter could be optimized via a trial-and-error strategy, however such a choice would require running a couple of iterations whilst monitoring the residual norm until a successful step-size is found. These additional gradient computations would impact the overall cost of the inversion. 

\begin{figure}
  \centering
  \includegraphics[width=0.5\textwidth]{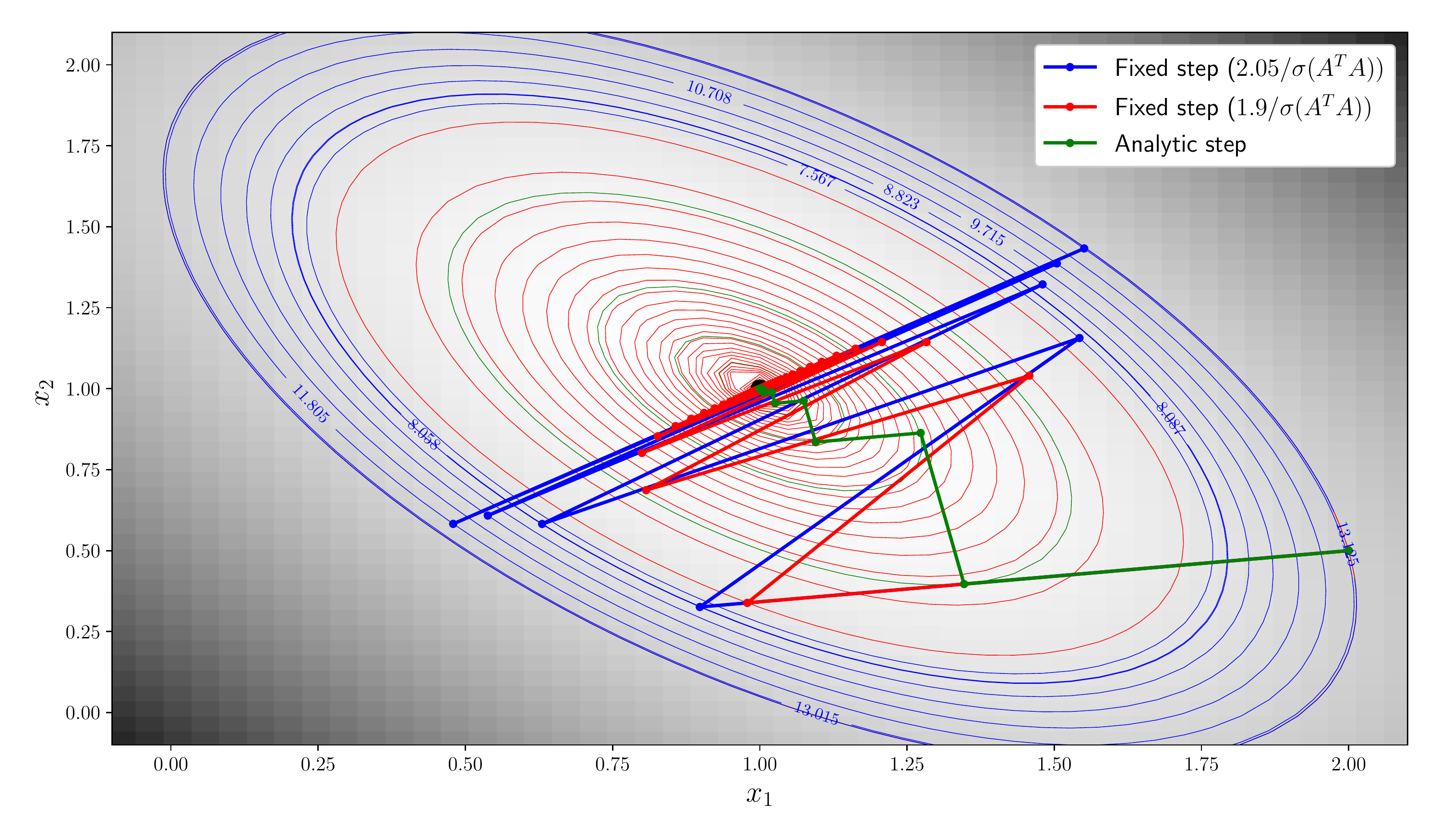}
  \caption{Convergence trajectories for Landweber iterations with step-size smaller (red) and larger (blue) than maximum allowed value, and steepest descent algorithm (green).}
  \label{fig:steepest}
\end{figure}

Alternatively, we propose to use a semi-heuristic procedure based on the well-known Landweber iteration \cite{landweber1951}, a variant of the gradient descent algorithm with fixed step-size. Simply put, given a linear inverse problem (e.g., $\textbf{x} = \underset{\mathbf{x}} {\mathrm{argmin}} \frac{1}{2}||\textbf{y} - \textbf{A} \textbf{x}||_2^2$), Landweber showed that convergence is guaranteed for the following choice of the step-size, $0<\alpha< 2/\sigma^2(\textbf{A})$. Here, $\sigma(\textbf{A})$ is largest singular value of the matrix (or operator) $\textbf{A}$. Such a condition can be easily verified with a numerical example:
\begin{equation}
\label{eq:landwebersystem}
\textbf{A}= 
\begin{bmatrix}
    1 & 4 \\
    1 & 3 \\
    6 & 2
  \end{bmatrix}, \quad 
\textbf{x}= 
\begin{bmatrix}
    1 \\
    1
  \end{bmatrix}
\end{equation}
where $\sigma^2(\textbf{A})=53$. Fig. \ref{fig:steepest} shows the convergence behaviour of the Landweber algorithm for two choices of step-size: $\alpha=2.05\sigma^2(\textbf{A})$ (blue line) and $\alpha=1.9\sigma^2(\textbf{A})$ (red line). It is clear that convergence to the solution cannot be achieved when the above condition is not met, even when $\alpha$ is chosen to be only $2.5\%$ larger than the maximum allowed value. Finally, the Landweber iteration is also compared to the steepest descent algorithm (green line), where the step-size varies at each iteration and is chosen to minimize the cost function along the selected gradient direction. Whilst this selection criterion leads to a much faster convergence, it cannot be easily extended to stochastic gradient descent algorithms.

\begin{figure}
  \centering
  \includegraphics[width=\textwidth]{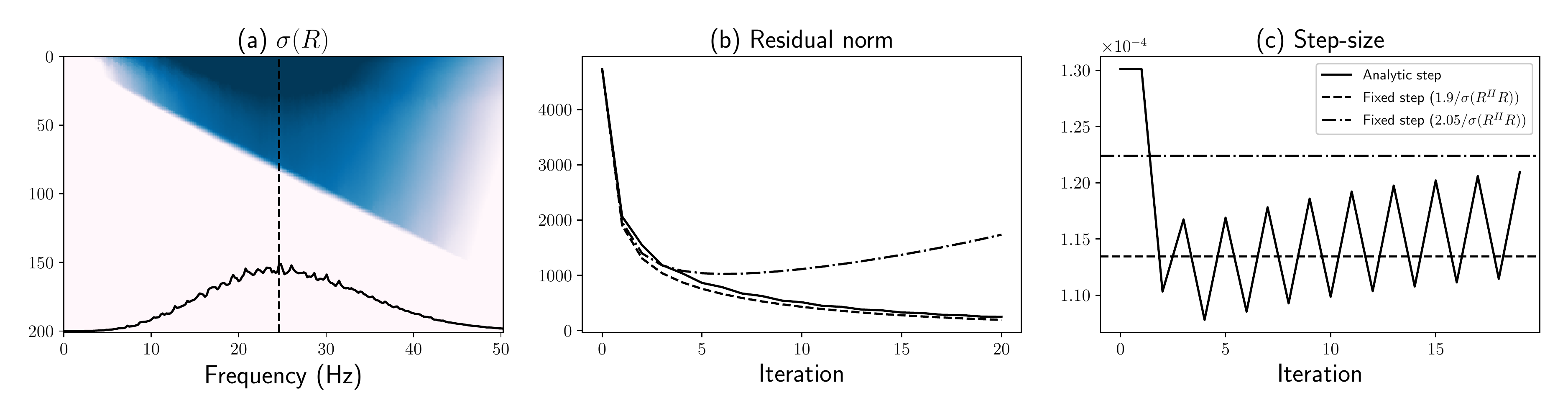}
  \caption{a) Eigenvalue spectrum (image) and average frequency spectrum (solid line) for the Synthetic OBC down-going wavefield in frequency domain. A vertical dashed line indicates the frequency associated with the peak of the amplitude spectrum (and the maximum eigenvalue). b) Residual norm and c) step-size as function of iterations for the steepest descent algorithm (solid), the Landweber algorithm with step-size smaller (dashed), and larger (dashed and dotted) than maximum allowed value.}
  \label{fig:steepestobc}
\end{figure}

Moving back to the MDD problem, a question arises about how to compute the largest singular value of the Multi-Dimensional Convolution operator. Whilst estimating it by means of matrix-free power iteration methods is very expensive as it requires the evaluation of several forward and adjoint passes of the operator, \cite{luiken2020} showed that the singular values of the operator are equivalent to those of the kernel (i.e., frequency-domain down-going wavefield). Furthermore, as shown in Fig. \ref{fig:steepestobc}a, 
the largest singular value of the operator is associated with the frequency matrix with strongest contribution to the data; this can be easily estimated by looking at the average spectrum of down-going wavefield (black curve at the bottom of Fig. \ref{fig:steepestobc}a). Fig. \ref{fig:steepestobc}b shows the residual norm as a function of iterations when solving equation \ref{eq:mddtinv} with the Landweber iteration using a step-size equal to $1.9 / \sigma(\textbf{P}_t^{+H} \textbf{P}_t)$ and $2.05 / \sigma (\textbf{P}_t^{+H} \textbf{P}_t)$ as well as the analytical optimal step size.

Heuristically, we found that for stochastic MDD the choice of the step size must be made more conservatively due to the fact that inexact gradients are used at each step; in other words, by using a step-size of $40-60\%$ of the theoretical upper limit for the Landweber iteration we are able to consistently produce stable solutions across the various numerical examples presented in this paper.

\section*{Acknowledgment}
The authors thank KAUST for supporting this research. All numerical examples can be accessed at https://github.com/DIG-Kaust/MDD-StochasticSolvers. Moreover the synthethic and real Volve datasets can be accessed at https://github.com/DIG-Kaust/VolveSynthetic and https://data.equinor.com. We are grateful to Equinor and partners for releasing the Volve dataset. MR thanks Claire Birnie (KAUST) for insighful discussions.

\bibliographystyle{unsrt}  
\bibliography{references}

\end{document}